\documentclass[twocolumn]{aastex631}

\shorttitle{Instrument Model Characterization through Wide Scale Analysis}
\shortauthors{Huber-Feely et al.}
\graphicspath{{./}{figures/}}
\begin{document}

\title{Characterization of an Instrument Model for Exoplanet Transit Spectrum Estimation through Wide Scale Analysis on HST Data}

\correspondingauthor{Noah Huber-Feely}
\email{nh2567@columbia.edu}

\author[0000-0003-1906-0030]{Noah Huber-Feely}
\affiliation{Jet Propulsion Laboratory, California Institute of Technology, 4800 Oak Grove Drive, Pasadena, California 91109, USA}
\affiliation{Columbia University, New York, NY 10027, USA}

\author[0000-0002-0919-4468]{Mark R. Swain}
\affiliation{Jet Propulsion Laboratory, California Institute of Technology, 4800 Oak Grove Drive, Pasadena, California 91109, USA}

\author[0000-0002-7402-7797]{Gael Roudier}
\affiliation{Jet Propulsion Laboratory, California Institute of Technology, 4800 Oak Grove Drive, Pasadena, California 91109, USA}

\author[0000-0002-4006-6755]{Raissa Estrela}
\affiliation{Jet Propulsion Laboratory, California Institute of Technology, 4800 Oak Grove Drive, Pasadena, California 91109, USA}
\affiliation{Center for Radio Astronomy and Astrophysics Mackenzie (CRAAM), Mackenzie Presbyterian University, Rua da Consolacao, 896, SaoPaulo, Brazil}



\begin{abstract}

Instrument models (IMs) enable the reduction of systematic error in transit spectroscopy light curve data, but, since the model formulation can influence the estimation of science model parameters, characterization of the instrument model effects is crucial to the interpretation of the reduced data. We analyze a simple instrument model and assess its validity and performance across Hubble WFC3 and STIS instruments. Over a large, $n=63$, sample of observed targets, an MCMC sampler computes the parent distribution of each instrument model parameter. Possible parent distribution functions are then fit and tested against the empirical IM distribution. Correlation and other analyses are then performed to find IM relationships. The model is shown to perform well across the 2 instruments and 3 filters analyzed and, further, the Student's t-distribution is shown to closely fit the empirical parent distribution of IM parameters and the Gaussian is shown to poorly model the observed distribution. This parent distribution can be used in the MCMC prior fitting and demonstrates IM consistency for wide scale atmospheric analysis using this model. Finally, we propose a simple metric based on light curve residuals to determine model performance, and we demonstrate its ability to determine whether a derived spectrum under this IM is high quality and robust.

\end{abstract}



\section{Introduction}

Instrument models (IMs) are needed to marginalize systematic error in exoplanet transit data. These marginalization parameters are typically computed by determining the values that best fit the data. This fitting is needed because the IM parameters for HST have so far been unable to be fully predicted from the instrument and target characteristics alone. Although the physical conditions vary from one observation to another, if instrument noise originates from the same instrument effects and under an IM parametrization that suppresses the specifics of individual cases, the best fit instrument model parameters are expected to follow a common empirically based set of parameter priors across the targets of a particular instrument. For Bayesian fitting approaches, quantifying and modeling these priors is important for robust and non-degenerate fits. As current IM models are primarily phenomenoligical rather than physically based, setting priors and uncertainties is challenging, and we thus attempt to explore an empirical approach to this problem.
   
This study computed high resolution spectra from the filters, targeting resolutions near the dispersion limited spectrum resolution. We defined this resolution to be a pixel binning at a spatial width on the detector near the point spread function full width at half maximum (PSF FWHM). For WFC3 G141, this resolution is the same as the native detector pixel resolution, and for the STIS filters we were significantly above the FWHM to mitigate noise effects, but still at high resolution. This was done to fully explore error effects, and to analyze the capabilities of a native resolution IM model. By using the full resolution spectrum, atmospheric analyses can employ even more fine-grained spectral features, and bad spectral channels can be identified and can be excluded to prevent any resulting biasing effects.

Additionally, our analysis aims to test whether there are fundamental differences found when we apply the same IM to multiple instruments, which in our study is between the HST sensors WFC3 and STIS. To characterize this IM, we need to quantify its performance in a multi-instrument, large source catalog environment. We do this by assessing the IM on 63 targets spanning these 2 instruments.
   
   Previous work such as \citet{wakeford16} have determined ways to marginalize instrument error, but wide scale distributional analysis has so far not been done on the derived parameter values, as opposed to simply the quality of fit. As instrument marginalization could bias derived spectrum, analyzing the derived parameters is important to characterize the IM and explore the validity of these instrument marginalizations.
   
   The formulation of our proposed IM has been shown to have connections to physical parameters. Specifically temperature fluctuations and buffer dumps have been identified as correlated, while one parameter, the visit long light curve slope, has so far been unable to be correlated to any instrument characteristic, see \citet{wakeford16}. The analysis in this paper uses a different approach, namely exploring the purely distributional shape and behavior of these parameters, rather than physical explanations. This focus was chosen as these other connections, while important have not fully explained the parameters.
   
   Further, several papers such as \citet{gibson14} have analyzed how closely instrument model systematics derive the correct synthetic spectra for generated light curves. Our paper's analysis, while reserved to a single instrument model, attempts to answer similar questions to this paper while using observed data rather than synthetic observations. We achieve this by analyzing how the marginalization of individual targets compares to the marginalization across all 63 targets under consideration in this paper. By analyzing this parent distribution, we can verify if IM parameters are correlated with science parameters in such a way that they would induce a false science signal, or if any unexpected or interesting correlations exist. Further, we can determine if there is evidence for an IM parent distribution and can thus propose a robust way to use this empirical distribution for the setting of MCMC priors and assessment of our IM.
   
   Recent papers such as \citet{laginja20} have explored and provided methods for marginalizing systematics across a grid of different IM formulations. This grid marginalization approach can provide many benefits and practical applications. Our paper instead focuses on a single IM formulation so that it can explore greater depth of analysis. More IM formulations, such as ones incorporating polynomial marginalization, could be explored under our paper's methods in the future, and this is a ripe area for future research.
   
   Additionally, recent work has continued to improve Bayesian setups using updated prior values. One primary example is the paper \citet{taaki20} which detected transits in the Kepler dataset which, although using a different setup than ours and focusing on a different space telescope dataset, used the empirical distribution to construct a best fit prior for its parameters. We expand upon this type of work by applying this parent distribution analysis to both the HST infrared grism G141 and STIS CCD sensors G430L and G750L.
   
   Finally, to determine the quality of improvements and modifications suggested by our validity analysis, we use a metric with units of estimated shot noise. This allows us to compare our performance to the estimated theoretical best performance while also allowing for general comparison of relative performance across instrument types.

\section{Methods}

\subsection{Data Reduction}

The reduction process from observed data to the unmarginalized light curves, while an important part of light curve analysis, does not make up the useful contribution of this work, so is not covered in great detail in this paper and rather uses the same methods employed and described in \cite{swain21}. All transit data used in analysis was retrieved from the MAST archive .ima data products \citep{dressel17}. As is commonly done in the literature \citep{kreidberg14}, the first orbit of each visit was not used in fitting or analysis due to its ramp effects behaving differently from the rest of the orbits.

The data for HST WFC3 G141 is reduced at the native pixel resolution rather than the four pixel averaging typically done, as seen in \citet{deming13}, \citet{kreidberg14} and \citet{tsiaras16}. For WFC3, the PSF FWHM ranges from 1.019 at 1.1 microns to 1.219 at 1.7 microns \citep{dressel17}. Thus, each pixel contains at most 76\% of the 1-D PSF for the given wavelength. Therefore, neighboring spectral bins, each of which contains data for a single pixel, will share some noise or signal across them. However, the dominant noise sources for G141 are pixel based such as flat field error, time dependent gains errors and amplifier drifts. Thus, given the relatively low SNR of the underlying data, this pure photon noise error correlation is not expected to bias the results. We thus chose to use this full resolution rather than halving or even greater reductions of resolution that binning would require. However, in other SNR regimes, such as for the James Webb Space Telescope (JWST), the amount and the effect of correlated noise would be an important aspect to study further for these kinds of analyses.

G430L and G750L are also reduced but at lower pixel resolution than G141 due to their wider PSF FWHM and higher SNR and this is done as described in \citet{Estrela21}. The spectral resolution (FWHM) of G430L is approximately 1.31 pixels and G750L is 1.46 pixels \citep{branton17}. We average 3 pixels for G430L and 7 pixels for G750L, thus ensuring the spectral channels are independent in terms of photon arrival, while still providing a high resolution spectrum for analysis.

The wavelength dependence of the systematics is an important object of study. By using the native or near native spectral resolution of the instrument, we can gain a more precise understanding of this relationship.

\subsection{EXCALIBUR Pipeline}

The EXoplanet CALIbration and Bayesian Unified Retrieval (EXCALIBUR) pipeline at JPL provided the computational framework for the project \citep{swain21}. Specifically, it is a pipeline that implements a method for uniform spectrum estimation for comparative exoplanetology. The specific architecture, while making computation easier, does not influence the spectrum estimation itself.

\subsection{Instrument Model Definition}

Our instrument model derives five parameters for each visit, which are then applied across both in-transit and out-of-transit data. To fit the light curve averaged across all spectral channels, what we call the whitelight light curve, we used the following instrument model definition:

\begin{eqnarray}
\nonumber
M(\mathbf{t})=M_{0}(\mathbf{t})[v_s (\mathbf{t}-T_\mu) + 1][o_s (\mathbf{t}-t_\mu) + o_i]\\
\bigg[1-\exp{\bigg(-\frac{(\mathbf{t}-t_0) + 10^{o_d}}{10^{o_\tau}}\bigg)}\bigg]
\end{eqnarray}

Here $M(\mathbf{t})$ is a vector of all the whitelight observations in the light curve, and $M_{0}(\mathbf{t})$ is the science model. $T_\mu$ is the mean of $\mathbf{t}$ (i.e. the center of the transit). $t_{\rm \mu}$ is set per orbit and corresponds to the mean of the time values in $\mathbf{t}$ computed for each particular orbit. $t_0$ is also per orbit and corresponds to the first time value in the corresponding orbit. $v_{\rm s}$, $o_{\rm s}$ and $o_{\rm i}$ are all estimated per visit and no parameter estimates are computed per orbit. Finally, $o_{\rm d}$ and $o_{\rm \tau}$ are called the exponential breath parameters and are applied over each orbit but their value is uniform across all orbits in a specific visit.

For spectrum fitting, the same IM formulation was applied, but the exponential breathing parameter fitted values from the whitelight fit were used. To represent this, all wavelength dependent parameters now have the $\lambda$ subscript and all others use the whitelight fitted parameters, but the parameter meanings all remain the same.

\begin{eqnarray}
\nonumber
M_{\lambda}(\mathbf{t})= M_{0,\lambda}(\mathbf{t})&[v_{s,\lambda} (\mathbf{t}-T_\mu) + 1][o_{s,\lambda} (\mathbf{t}-t_\mu) + o_{i,\lambda}] \\
&\bigg[1-\exp{\bigg(-\frac{(\mathbf{t}-t_0) + 10^{o_d}}{10^{o_\tau}}\bigg)}\bigg]
\end{eqnarray}

For ease of display and explanation, mentions of vslope or visit slope in the paper will refer to the $v_{\rm s}$ and $v_{\rm s,\lambda}$ parameters. Similarly, oslope or orbit slope will refer to $o_{\rm s}$ and oitcp or orbit intercept will refer to $o_{\rm i}$ as well as their analogous wavelength dependent equivalents.

\begin{figure}
   \centering
   \includegraphics[width=\linewidth]{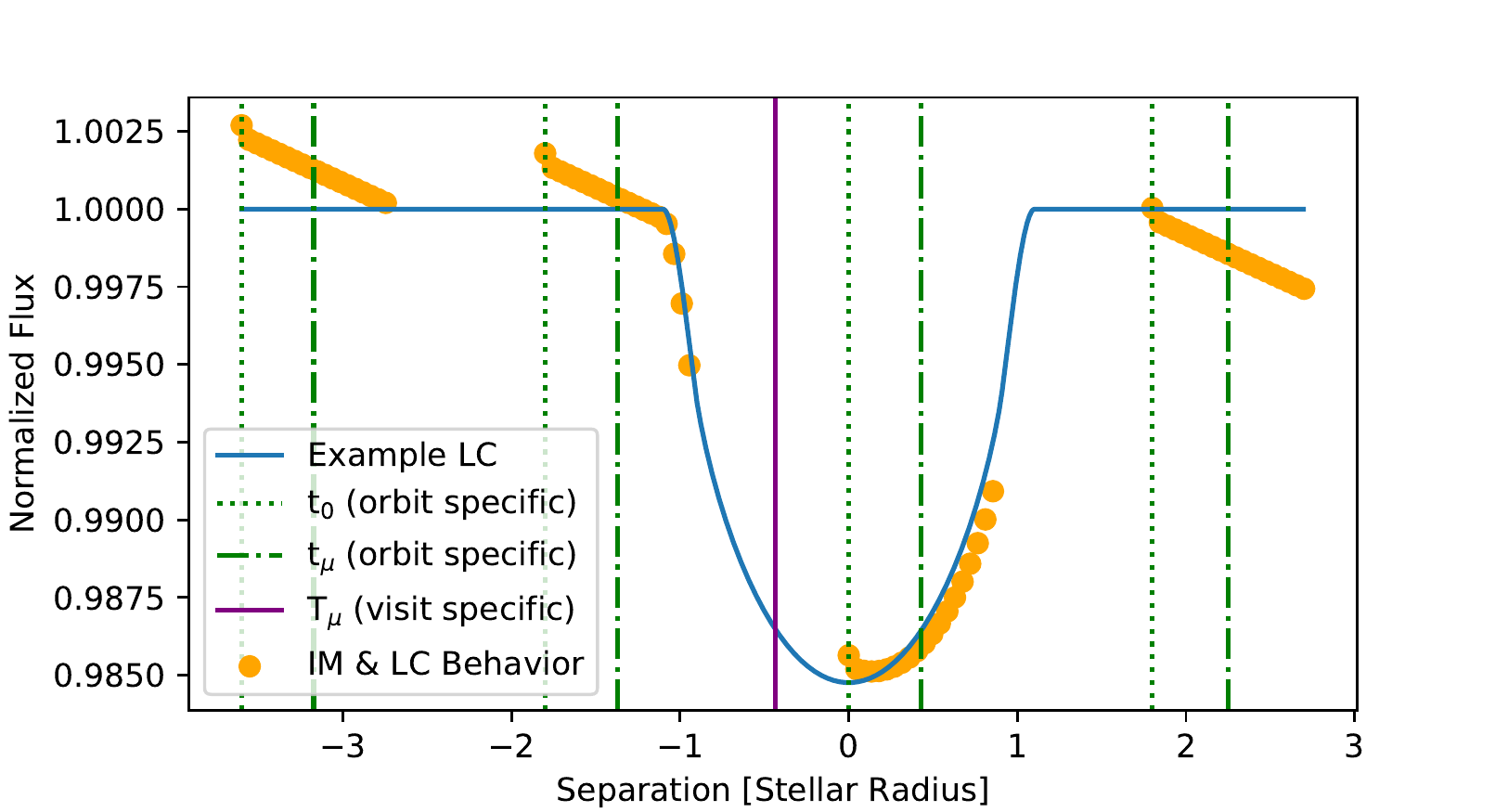}
   \caption{Example behavior of the IM on a synthetic light curve. The placement and balance of the timing parameters in the formulation is overlayed. This figure represents just one set of potential IM parameters, and the synthetic parameters chosen were selected for their clarity of displaying IM effects.}
              \label{ExampleIMResult}%
    \end{figure}
    
Fig.~\ref{ExampleIMResult} illustrates a synthetic example of the IM behavior with all other random effects such as photon noise removed to enhance interpretability. The placement of the timing parameters is visually demonstrated in the figure. \citet{wakeford16} explores and visualizes the same linear and exponential effects discussed in this section through empirical data.

\subsection{Instrument Model Discussion}

The key features of this IM formulation lie in how the timing data is defined and how the normalization, i.e. oitcp, term is integrated. As seen in papers such as \citet{tsiaras16}, the normalization term is often defined outside of the orbital slope term. This results in a mathematical simplification from $o_{i,\lambda}[o_{s,\lambda} (\mathbf{t}-t_\mu) + 1]$ to $[o_{i,\lambda} \cdot o_{s,\lambda} (\mathbf{t}-t_\mu) + o_{i,\lambda}]$. Thus, the slope term $o_{s,\lambda}$'s value is proportionally dependent upon the value of the normalization term, $o_{i,\lambda}$. In our formulation, we prevent this by expressing the two terms by $[o_{s,\lambda} (\mathbf{t}-t_\mu) + o_{i,\lambda}]$. This allows us to explore the slope effects separate from the normalization effects.

Further, by defining the orbital slope term as $o_{s,\lambda} (\mathbf{t}-t_\mu)$, the slope is defined relative to the center of each orbit. By doing so, we ensure that $o_{i,\lambda}$ is not dependent upon the slope term as it would be if the timing data was centered on the start time of each orbit. This is critical for valid analysis of the parameters as it helps to separate the two effects, while being equally expressive and equivalent to this other approach. An analogous argument holds for the definition of the visit long slope by $(\mathbf{t}-T_\mu)$ which helps mitigate any dependent effects between the visit slope and the normalization term.

Physical and empirical justifications for these linear effects for WFC3 are explored in \citet{wakeford16}. Two linear terms and a breathing effect were determined to be empirically justified for the IM. Further, the orbit slope has a fixed intercept such that a zero slope results in no effects from this linear term. The visit slope can thus be interpreted as enforcing a visit wide linear intercept trend, rather than each orbit having arbitrary intercepts.

The whitelight fit breathing parameters were reused for the spectrum fit rather than performing an additional per wavelength breathing fitting. This was chosen because the breathing effects have been linked to thermal effects as HST moves into and out of Earth's shadow. Thus there is no a priori justification for wavelength variation of this effect, and, further, there was no apparent empirical support. Conversely, the slope terms are empirically observed to vary by wavelength and, as will be discussed, this fit variation does not show indicators of overfitting, thus empirically supporting the wavelength dependent formulation for the linear terms.

\subsection{Instrument Model Fitting}

Under the above instrument model definition, a Metropolis-Hastings Markov Chain Monte Carlo (MCMC) posterior estimation was performed using the PyMC3 package \citep{salvatier16} and model parameters were selected by computing the median of the sampled parameter posterior. The exponential breathing parameters were fit using a $\chi^2$ minimization as will be discussed in the next paragraph. The MCMC fitting was done using 4 walkers, a chain length of 500,000 and a burn-in of 30\%-50\% of the chain. For initially determining this parameter configuration, the Gelman-Rubin diagnostic included in PyMC3 was used to assess convergence.

For the per channel spectrum light curves, the exponential breathing parameter from the whitelight fit was used as it has been shown to have no significant variation between wavelengths as noted by multiple authors including \citet{tsiaras16}. This fitting process was done through a $\chi^2$ minimization for these exponential breathing terms over the out of transit data. We used the lmfit library \citep{newville_matthew_2014} and found the included Conjugate-Gradient method to provide a fast and robust optimization. The rest of the parameters were then independently refit to each wavelength channels light curve data by MCMC.

The MCMC priors for initial analysis were set to be very wide and non-informative so as to not influence the results of the analysis. The MCMC prior for the orbital intercept parameter was set with a mean of 1.0 and a standard deviation equal to the standard deviation of the out of transit observations. The prior for both the orbit long and visit long slope terms were set to a mean of 0 and a very wide standard deviation. Specifically, this standard deviation represented the expectation that the slope would not exceed the transit depth in at most 68\% of cases.

\subsection{Distribution Analysis Methods}

    To analyze the IM parameter behaviour, the fit parameters and their MCMC computed posterior trace were extracted from the EXCALIBUR pipeline. For visual analysis, histograms were displayed, with multiple histograms plotted together when segmenting by characteristics to detect distributional differences between population sections. To test distributional fits, the Kolmogorov--Smirnov (KS) test was applied between the probability density function of the distribution and the empirical distribution of the parameters. The 1-sample t-test was applied when testing for mean significance.

Further, we applied a visual analysis method to visualize the range of IM effects expected under the estimated priors. Specifically, we took the LC model for one target, and then implemented a function to apply the linear IM effects to this LC. Finally, we sampled a large number of parameter sets from the estimated priors and then computed the LC with these IM effects applied. We then plotted the original LC model and several quantiles of the LCs with sampled linear IM effects applied. This thus allowed us to visually observe the range of combined expected effects under the estimated IM priors for visit slope, orbit slope and orbit intercept. Further, the systematics corrected light curve data was plotted so that the magnitude of the IM error effect could be compared to the magnitude of the LC model residuals.

\subsection{Residual Standard Deviation Analysis}

This papers analysis uses metrics derived from the light curve residual standard deviation. For a given spectrum, each wavelength channel has a corresponding light curve which has been fitted to an IM. The residuals of this are the difference between the observed data and the fitted IM and light curve model. The residual standard deviation is thus defined to be the standard deviation of these light curve residuals. This can be interpreted as the scatter of observations around the model. Further, shot noise was computed as described in \citet{batalha17}. We thus defined a metric of residual standard deviation per shot noise (RSDSN) as the ratio of the observed standard deviation around the model to the shot noise theoretical expectation. This metric thus captures the magnitude of variation relative to the theoretical best performance with a perfect sensor. Previous work has similarly used the photon-limited precision to compare the relative precision and fit of models, specifically \cite{stevenson19}.

\section{Exponential Breath Parameters}

As discussed in the Methods section, the exponential breathing parameters correct for thermal effects resulting from HST passing into Earth's shadow. We parametrize the effect as $[1-\exp{(-\frac{(\mathbf{t}-t_0) + 10^{o_d}}{10^{o_\tau}})}]$ which is applied per orbit. $o_d$ parametrizes the delay of this breathing effect and $o_\tau$ parametrizes the effect magnitude. These are fit against the light curve data before any instrument systematics corrections, and done via a $\chi^2$ minimization as discussed in Methods.

These parameters have well defined upper and lower bounds based on the raw timing data without respect to sensor configuration. Both the delay and the $\tau$ parameter have a lower bound of the log of the minimum number of seconds between observations in that orbit. The delay parameter is upper bounded by the log of 5 times the duration of the seconds in the selected time, $\log(5(\max(\mathbf{t_o})-\min(\mathbf{t_o})))$, and the $\tau$ parameter is upper bounded by simply the log of the duration of the seconds in the orbit, $\log(\max(\mathbf{t_o})-\min(\mathbf{t_o}))$. The simple interpretation of these values is that the exponential behavior will occur over at least 2 points in the light curve and will occur during the orbit.

The previous instrument model fitting began a $\chi^2$ minimization with the parameter starting values set to the mean of the upper and lower bound described above. Through population analysis of the fitted exponential breath parameters, we discovered that we could employ a significantly better starting value.

\begin{figure}
   \centering
   \includegraphics[width=\linewidth]{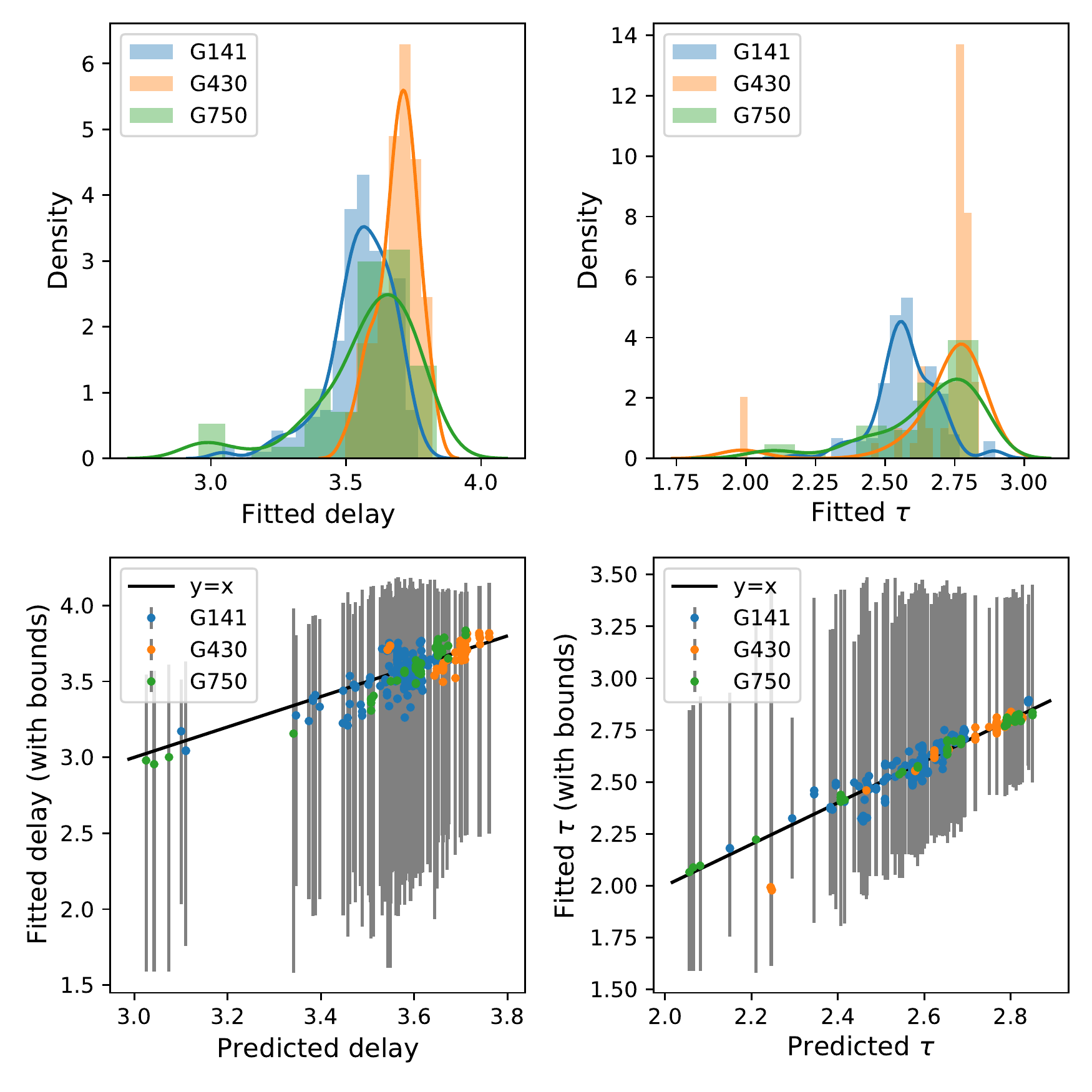}
   \caption{\textbf{Top:} Distribution of the delay and $\tau$ parameters across the three instruments. \textbf{Bottom:} Residuals of the previous naive starting value compared to the new single parameter starting value estimator. }
              \label{ExponentialBreathResiduals}%
    \end{figure}
    
To do so, we used a single parameter model to estimate the parameter value. Fig.~\ref{ExponentialBreathResiduals} shows the previous starting value residuals in orange and the new residuals of this start value model in blue. This model was fit across targets using two variables, $b_{up}$ for the upper bound and $b_{low}$ for the lower bound. The model $\alpha b_{up} + (1-\alpha) b_{low}$ was then fit for each target. Of all these fitted, $\alpha$ parameters the median was selected and used as the parameter value for all. This simple fitting process by using the median $\alpha$ across targets made the system more robust to outliers and skew as opposed to a least squares fitting process or other approach. The fitted $\alpha$ values for each filter and the two parameters are included in Table \ref{tab:ExponentialBreathParamFits}.

\begin{table}
\caption{Fitted values for the exponential breathing parameter relationship, $\alpha$.}
\label{tab:ExponentialBreathParamFits}      % is used to refer this table in the text
\centering                          % used for centering table
\begin{tabular}{c c c}        % centered columns (4 columns)
\hline\hline                 % inserts double horizontal lines
Filter & $o_d$ & $o_\tau$ \\    % table heading 
\hline                        % inserts single horizontal line
   G141 & 0.722 & 0.335\\
   G430L & 0.764 & 0.346\\
   G750L & 0.734 & 0.371\\
\hline                                   %inserts single line
\end{tabular}
\end{table}

Under this new fitted start value, the same $\chi^2$ optimization was performed and the resulting parameter estimates were analyzed. We discovered that the optimizer reached its stopping criteria very close to the newly derived starting value. Further analysis determined that this stopping criteria was met before the true optimum, but the fit quality difference between the stopped position and the true optimum was determined to be minimal with a $\chi^2$ improvement of under $5\%$. This demonstrated that this simple relationship captured most of the optimization. The rest of the optimization, while improving the fit some, did not result in dramatic improvements and thus simply using the determined relationship without data fitting could be satisfactory for some analysis.
    
\section{Parent Distribution Analysis}

We will now analyze the distributional characteristics of the linear IM terms. This analysis will explore the overall distribution across all targets, rather than the distribution conditioned on a specific target. We refer to this distribution as the parent distribution as it characterizes the distribution over all individual target priors, and is thus a hyper prior modeling the parameter behavior unconditioned on any given target. This will thus model and characterize the per filter/instrument prior distribution of these IM parameters. This can both provide further insights into the nature and dynamics of the IM as well as providing an empirical approach to selecting priors for an MCMC fitting.

\subsection{Whitelight Prior Analysis}

In our pipeline, we extract all the fitted whitelight light curve IM parameters for HST WFC3 G141 targets in scan mode, HST STIS G430L in stare mode, and HST STIS G750L in stare mode.

\begin{figure*}
   \centering
   \includegraphics[width=0.9\textwidth]{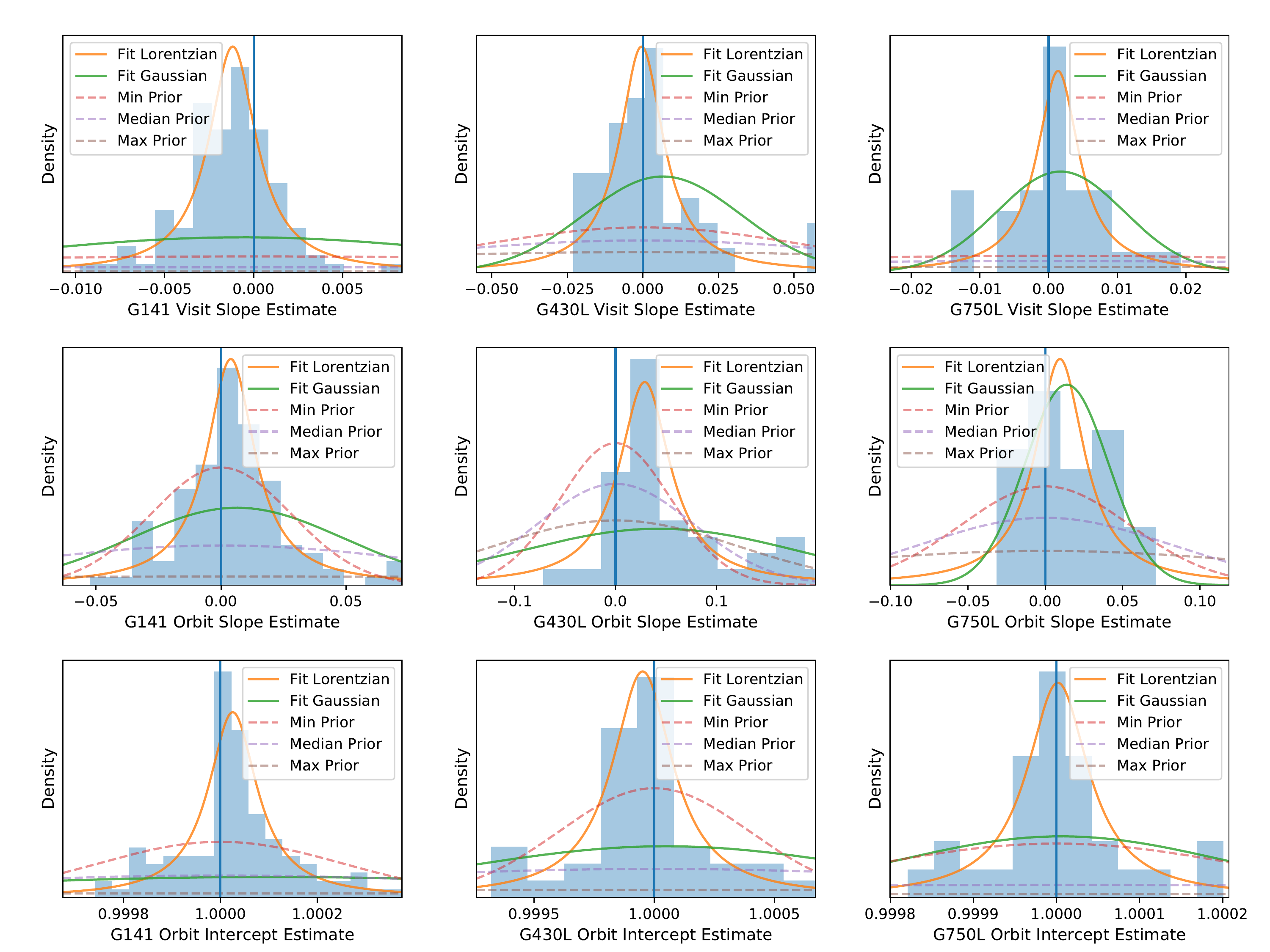}
   \caption{Empirical whitelight IM parameter distributions across instruments compared to the maximum likelihood Gaussian (green) and Lorentzian (yellow) distribution fits. The x-axis range for each parameter has been set so that the central distribution can be clearly seen. This results in outliers being cut off. The behavior of these outliers will be discussed in the text. The distribution of previously used priors is represented by the dashed lines. This prior distribution, as evidenced by the width of the median prior, is consistently wider than the empirical distribution for the visit slope and orbit intercept parameters. Additionally, the orbit slope priors also show a too wide fit for G141 and G750L with G430L orbit slope the only potential exception to this.}
              \label{WhitelightFittedParentDistr}%
    \end{figure*}

    In Fig.~\ref{WhitelightFittedParentDistr}, we show the empirical estimated IM parameter distribution, represented by the blue histogram, relative to the previously set prior values, represented by the dashed lines. The exact value these were set to is described in the methods section but they are all based on orbital and light curve properties and intended to be only mildly informative. The plot shows that the priors consistently overestimate the variance of the empirical distribution for the visit slope and orbital intercept terms, with the orbital slope minimum prior agreeing tightly in some cases, but underestimating the spread of the data as seen in G430L orbit slope. Wide priors or ones that underestimate the data spread can result in sampling inefficiencies and for targets without a strong data signal can lead to degenerate results. We thus turn our analysis to fitting and characterizing this empirical distribution.

Firstly, we considered the mean of these parameters. From Fig.~\ref{WhitelightFittedParentDistr}, we observed that the mean of the slope parameters was close to 0. As these slopes are computed after the exponential breathing parameter effects have been removed, this 0 centered behaviour of the orbit slope was consistent with expectation as these parameters marginalized linear behaviour beyond the exponential ramp effect which marginalizes some slope behaviour. The orbit intercept terms tightly fit the mean of 1, which aligns with a no effects model which is consistent with expectation as the fitted light curve should follow the calibrated out of transit centering closely.

To further analyze these distributions, we fit a maximum likelihood Gaussian to the empirical parent distribution as can be seen Fig.~\ref{WhitelightFittedParentDistr}. Under the Kolmogorov–Smirnov (KS) test, the fit of this distribution was rejected at the 5\% significance level. The fitted Gaussian is observed to be very wide. This is a result of the outlier parameters and wide tails causing the maximum likelihood fit to be significantly wider than the central distribution would suggest. This phenomomen of outlier parameters and wide tails resulting in a Gaussian fit being very broad was observed in \cite{taaki20} in their analysis of the Kepler dataset.

Our visual analysis thus indicated that for all parameters, this empirical parameter distribution was characterized by a strong central peak and significant outliers, indicating that a tighter central distribution and fatter tails than a Gaussian would provide a better fit. The Student's t-distribution of 1 degree of freedom, commonly called the Lorentzian, was selected as it possessed these qualities. The Student's t-distribution more generally has been used elsewhere in astrophysics for similar situations such as in \cite{kazemi13} where the Student's t-distribution was used for more robust calibration in a situation with frequent outliers. The Lorentzian distribution, by having fat tails, can better represent outliers while still tightly fitting a central mean. The Lorentzian distribution does however predict more outliers than observed in the empirical parameter distribution. This suggests that a slightly higher degree of freedom Student's t-distribution would offer an improved fit to the data as it would still impose the strong central mean but with reduced outlier behavior. However, given the small data size of the whitelight estimates, we used a Lorentzian for this part of the analysis as a fitted low degree of freedom Student's t-distribution would only offer marginal fit improvement given the small data size.

The KS test, for all the IM parameter distributions across every instrument, does not reject a maximum likelihood Lorentzian fit. For G141 with 59 targets, the distributional fit can have higher confidence than G430L and G750L with 27 and 29 exoplanet targets respectively. Positive confirmation that the samples follow a Lorentzian distribution would require an even larger sample size to substantively assert that it is the true underlying distribution of this whitelight fit. However, the KS test p-value exceeded 85\% for more than half of IM parameter distributions and failed to reject for the others, thus providing suggestive evidence that in the Bayesian prior context, this Lorentzian prior can be used without biasing the distribution.

We tested rerunning the MCMC with these new priors, and discovered that it slightly improved the light curve fit for some and had a decrease for others but there was not an apparent significant change in the resolved fit quality. This agrees well with a premise that the whitelight data signal is strong and thus the priors do not significantly affect the fitted estimates.

\subsection{Student's t-distribution Behaviour}

\begin{figure}
   \centering
   \includegraphics[width=\linewidth]{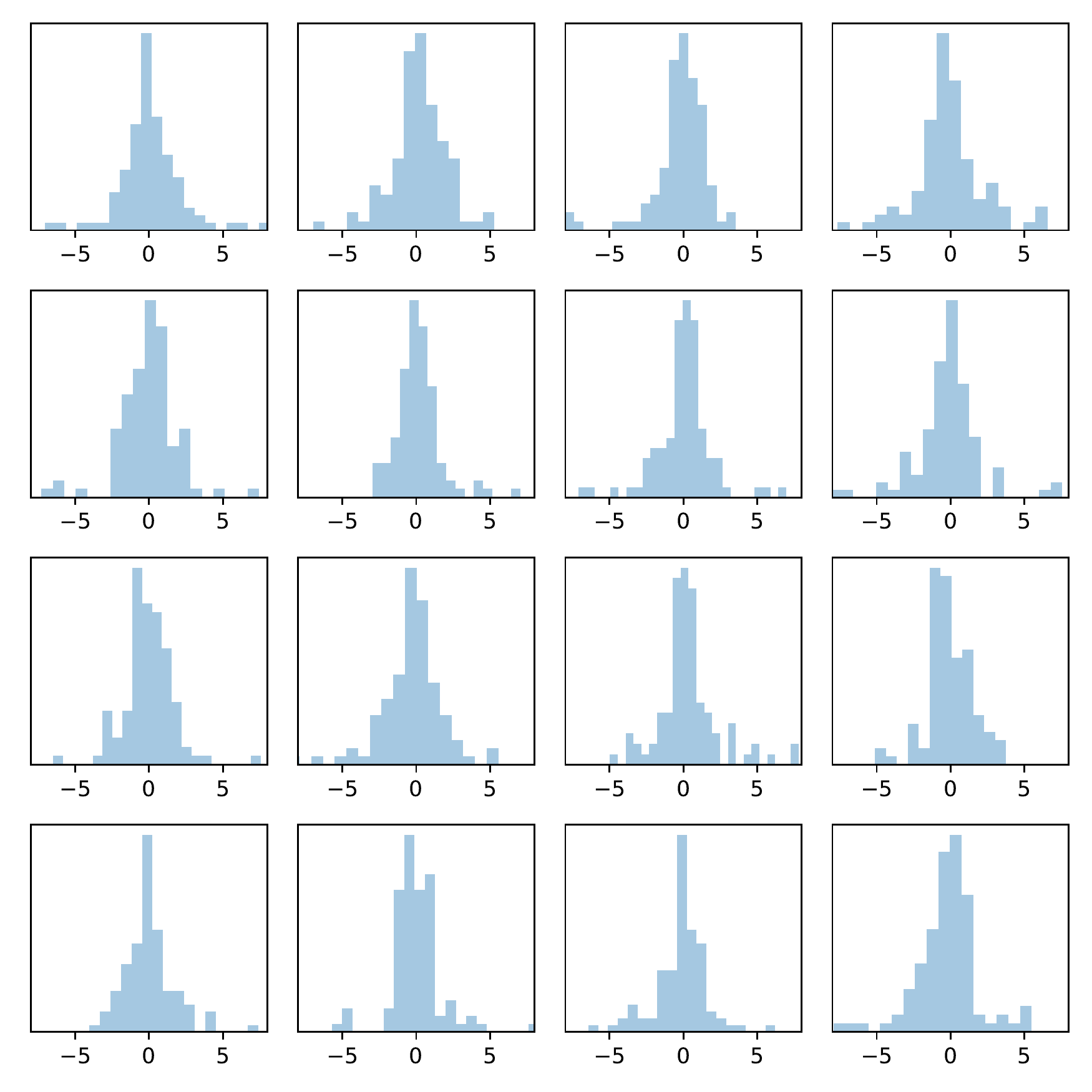}
   \caption{Histograms for 16 random samplings of a Student's t-distribution with 1.6 degrees of freedom at a sample size of 100. This plot highlights the range of possible behaviour seen under a low degree of freedom t-distribution assumption. The x-axis is fixed across the plots so some outlier points are cut off.}
              \label{T_Dist_Samples}%
    \end{figure}

As previously noted, the Lorentzian is a t-distribution of 1 degrees of freedom and, further, the Gaussian is a t-distribution of infinite degrees of freedom. The previous demonstration of the Lorentzian's potential fit quality and representativeness, does not rule out that a higher degree of freedom t-distribution might truly capture the distributional nature better. Fig.~\ref{T_Dist_Samples} is included to give the reader a sense of the range of behavior exhibited by a Student's t-distribution of low degrees of freedom. Some instantiations appear more Gaussian than others and the strength of the central peak and spread of outliers can differ significantly.  Further, the whitelight histograms have fewer samples than depicted in the Fig.~\ref{T_Dist_Samples} and thus even more variation is expected.

Thus, to gain further certainty in the possibility of a Student's t-distributed parent distribution, we need to significantly increase the sample size. We thus turn our attention to the per spectral channel IM estimates, which drastically increases the data size.

\subsection{Spectrum Prior Analysis}

The spectrum IM parameters, as they are fitted per wavelength, have many more parameter samples and thus analysis on them is significantly less data starved than analysis of whitelight parameters. Specifically, for G141, we now have more than 100 spectral channels per visit, each with independently fit parameters, and a similar number for G750L and G430L.

\begin{figure}
   \centering
   \includegraphics[width=\linewidth]{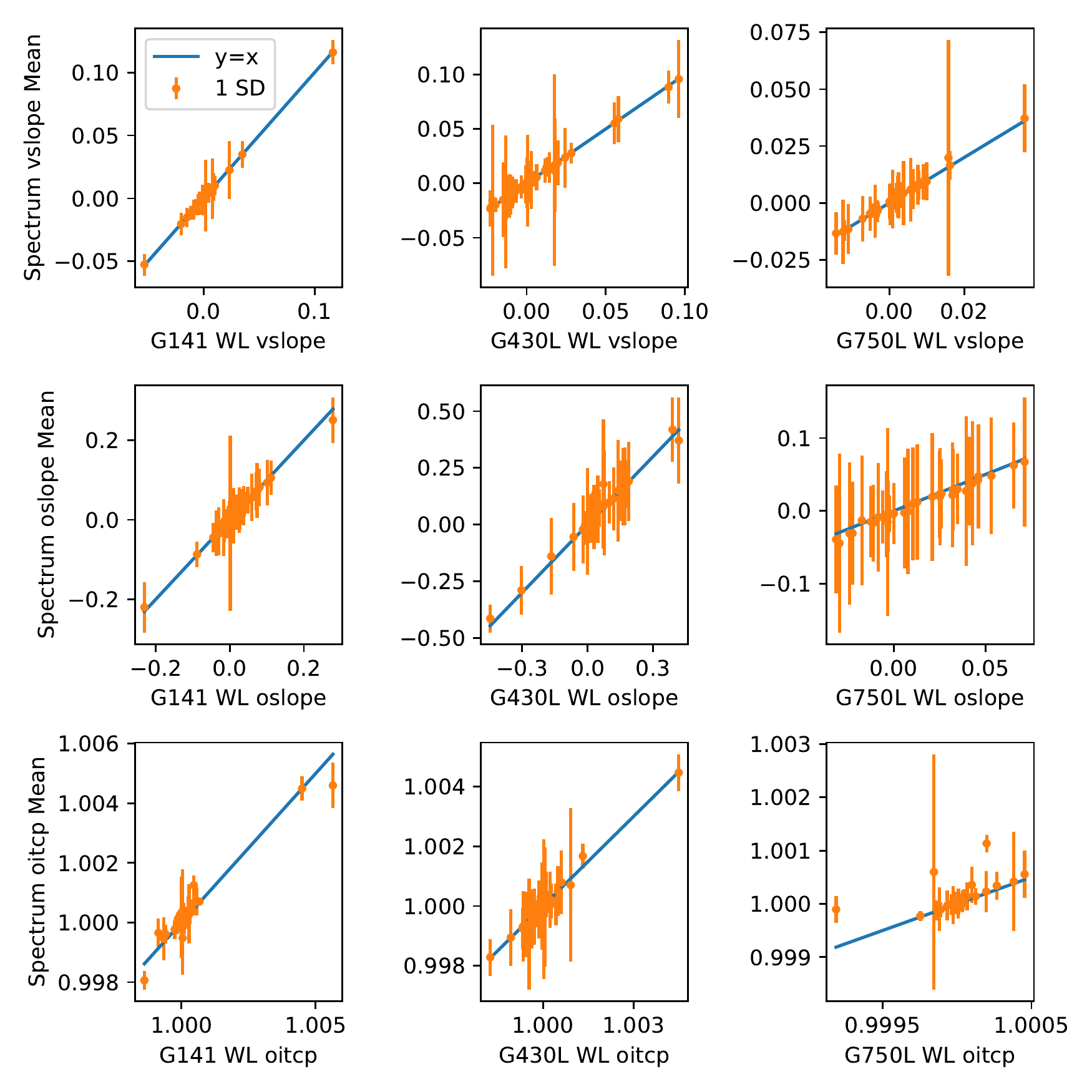}
   \caption{IM estimates for the whitelight (WL) light curves plotted against the mean of the spectrum IM estimates with the standard deviation of spectrum IM estimates for that target displayed as well. The blue line represents $y=x$ and thus if a point lies along it, this means that the spectrum value is equal to the whitelight value. From this, it can be seen that for almost all targets, the whitelight estimate is in close agreement to the average of the per target spectrum estimates. From this we can conclude that the spectrum estimates are not independent from the corresponding whitelight estimate.}
              \label{Spectrum_Whitelight_relation}%
    \end{figure}

However, the spectrum fit IM parameters for a single visit are not fully independent. Specifically, for each visit, we compared the average IM parameter across all spectral channels to the whitelight fit parameter for that same visit. We can see that the per spectrum fits agree closely with the whitelight fits for all visits and targets as shown in Fig.~\ref{Spectrum_Whitelight_relation}. This correlation between the whitelight parameter and the spectral average rejects the independent and identically distributed assumption of this sample, and thus the degrees of freedom of the sample is not calculable. Thus statistical tests for determining if the parent distribution of spectrum parameters fits a proposed prior distribution are not possible as the actual degrees of freedom is unknown.

\begin{figure*}
   \centering
   \includegraphics[width=0.9\textwidth]{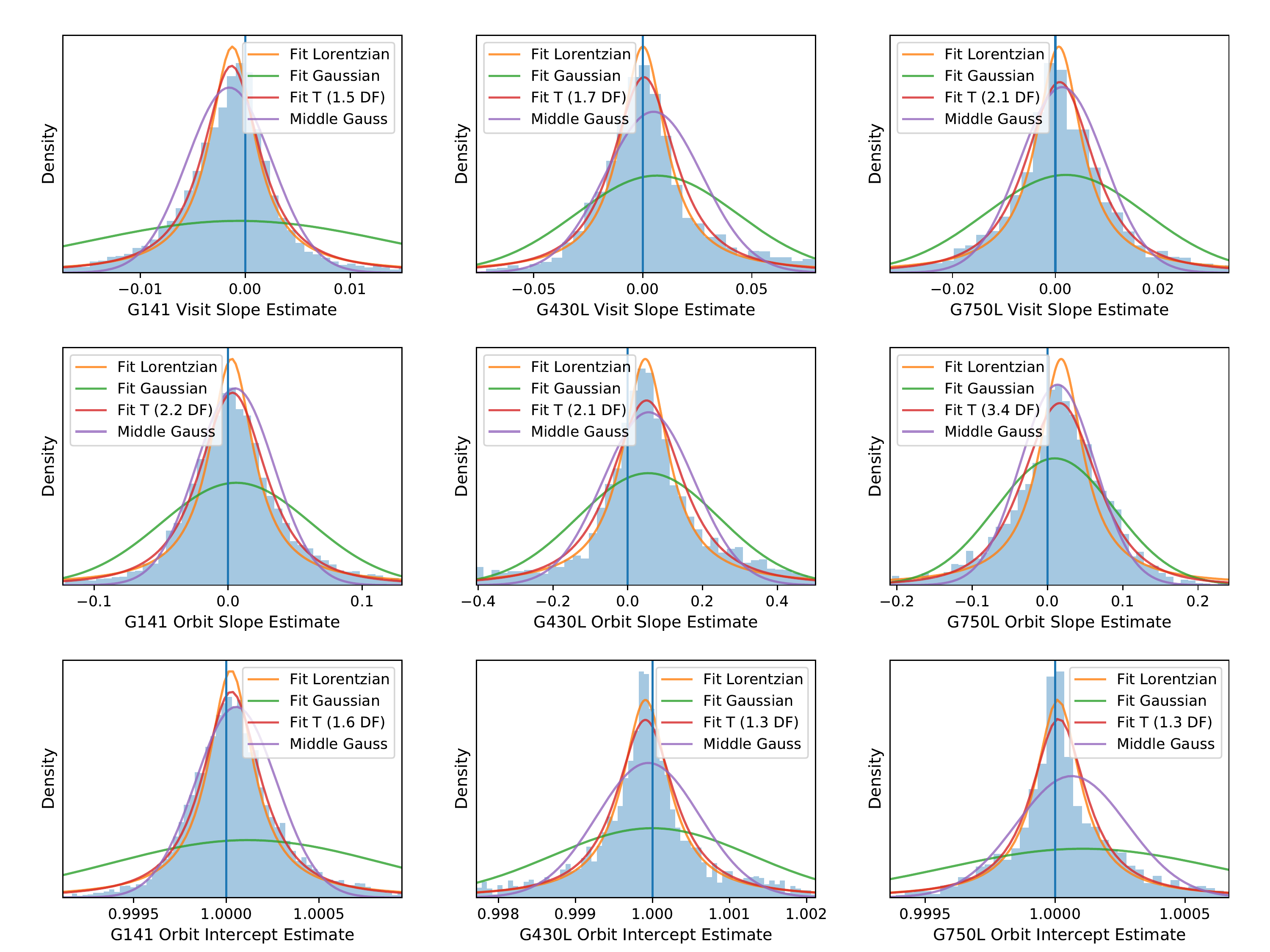}
   \caption{IM parameter distributions across instruments for per wavelength instrument model fits. The blue vertical line marks the center of the prior used in the initial MCMC fitting. The x-axis scale has been selected to visualize the central distribution and thus cuts off outliers, which is discussed in the text. A maximum likelihood fitted Gaussian, Lorentzian and Student's t-distribution are displayed. The Student's t-distribution is very close to the Lorentzian and provides the highest quality of fit. As a result of the outliers and empirical tail width, the fitted Gaussian is significantly wider than the central distribution suggests. Additionally, the Gaussian fitted only to the middle 90\% of the distribution significantly underestimates the tail probabilities and fits the central distribution worse than the Student's t.}
              \label{SpectrumFittedParentDistr}%
    \end{figure*}
    
\begin{figure}
    \centering
    \includegraphics[width=\linewidth]{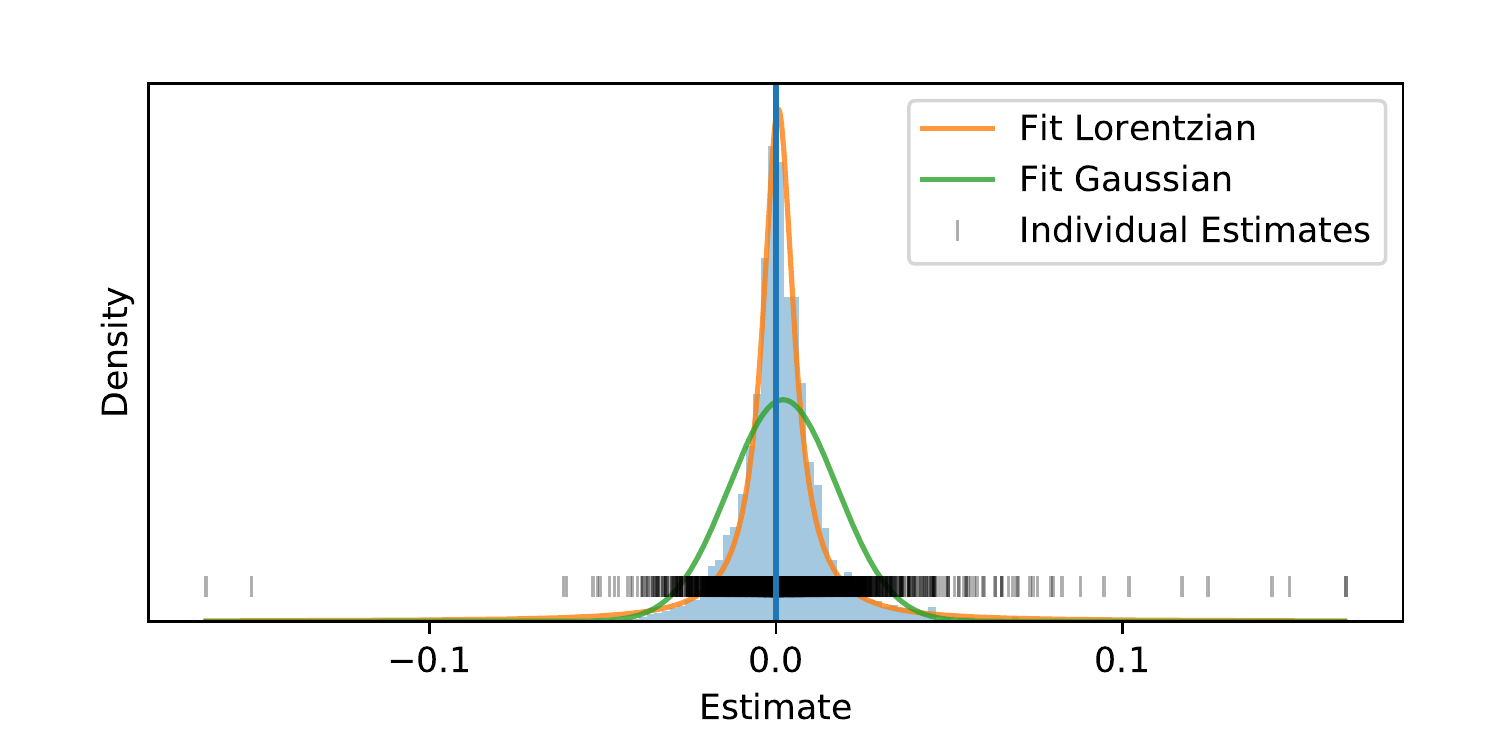}
    \caption{Non-truncated histogram of all fitted parameters for the G750L visit slope. Vertical tick marks are added for each parameter estimate to display both outliers and the central distribution.}
          \label{SpectrumDistrFullWide}%
\end{figure}
    
The empirical IM parameter distribution for the spectrum fits with all parameters binned together across all targets and visits is shown in Fig.~\ref{SpectrumFittedParentDistr}. Outliers, of which there were many, are cut off at the axis range. This was necessary to make the central distribution visible. Not shown are the previously used Gaussian priors in the MCMC, but these had equivalently poor quality of fit as shown previously in Fig.~\ref{WhitelightFittedParentDistr} for whitelight which suggests that tighter priors are justified for the distribution. A non-truncated distribution plot for the visit slope of G750L is displayed in Fig.~\ref{SpectrumDistrFullWide} to clarify the spread and outlier characteristics of the cut off data in Fig.~\ref{SpectrumFittedParentDistr}. G750L's outlier characteristics were a middle ground between the high spread and outliers of orbit intercept and the more centralized but still frequent outlier orbit slopes.

Using the same fitting approaches as in the whitelight section, Gaussian and Lorentzian distributions were fitted. Additionally, we now fit a Student's t-distribution with the degrees of freedom one of the optimized distribution terms. The analysis conclusions follow from the whitelight analysis, but visual inspection now increases confidence in the previous assertions of quality of fit. G141, G430L and G750L by visual inspection demonstrate well constrained Lorentzian behaviour with a Gaussian distribution too wide to be well fitting. The width of the Gaussian is significantly wider than the central peak in Fig.~\ref{SpectrumFittedParentDistr} would suggest and this is caused by the many outliers outside of the plot range. As noted for whitelight, this same phenomenon of outlier parameter induced widening of a fit Gaussian prior was observed in \citet{taaki20} for Kepler, and we conclude that a similar dynamic occurs for the HST slope systematics parameters. For illustration purposes, we also fitted a Gaussian to only the middle 90\% of the data. This Gaussian, termed middle Gauss in the legend, significantly underestimates the tail probabilities for all parameters, approaching 0 near the edges of the graph. For the outliers outside of the axis range, of which there were many, this Gaussian is of course even worse fitting and biasing from the true fit. Further, from visual observation it does a worse job tightly fitting the central tendency than the Student's t, as the Gaussian is not peaked enough to reflect the empirically observed distribution. A Gaussian, even one just fitted to the central tendency, thus does not capture the important properties of the empirical distribution and assigns a nearly 0 probability to commonly observed parameters.

For the the Student's t-distribution fit, we see that it closely agrees with the Lorentzian in most cases while providing an improved fit quality as is most evident in the G750L orbit slope plot in Fig.~\ref{SpectrumFittedParentDistr}.

The visit slope and orbit slope distributions appear to be roughly equivalent across G141, G430L and G750L with only mean shifting and spread differences between them. orbit intercept on the other hand shows very Lorentzian behavior for G141, but appears to show an even stronger central peak and wide tails for both G430L and G750L. This indicates a potential difference between WFC3 and STIS. However, the overall quality of fit of the t-distributions appears to be sufficient to well approximate orbit intercept across the 3 filters despite this distributional difference.

No distributional fit test could be correctly applied to the empirical IM parameter distributions as the samples did not meet the independent and identically distributed assumptions of the sample group due to the visit long dependence of the wavelength channels demonstrated in Fig.~\ref{Spectrum_Whitelight_relation}. We are thus left with a visual analysis as the statistically valid analysis means.

\begin{figure}
   \centering
    \includegraphics[width=\linewidth]{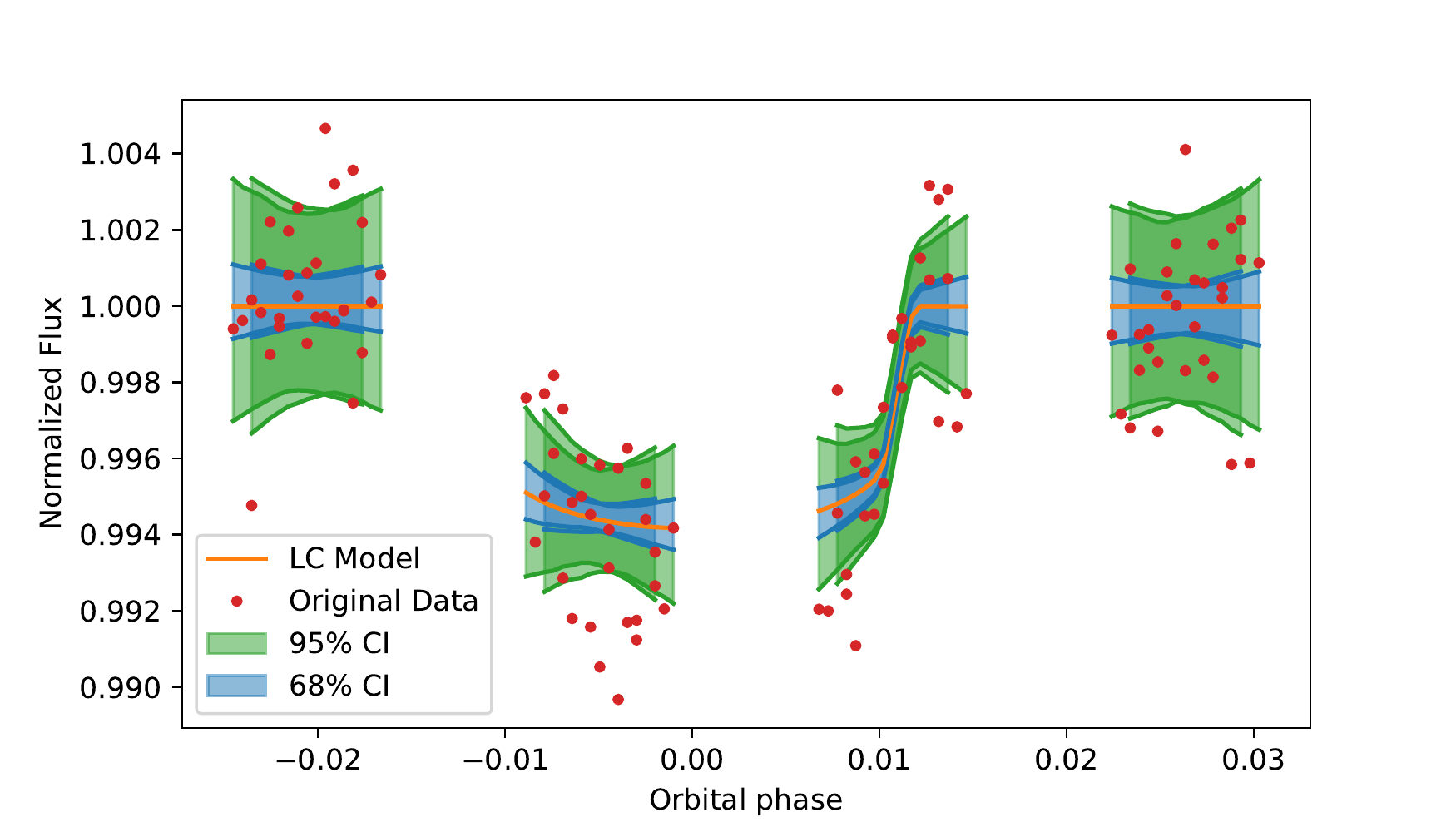}
   \caption{The distribution of forward modelled IM samples from the spectrum parent distribution for HAT-P-26 b at 1.137 microns for G141. The confidence intervals (CIs) demonstrate the magnitude of the effect the sampled systematics have on the light curve. Note that HAT-P-26 b has data from multiple visits and the overlapping CIs in the plot represent the estimates for the separate visits as each had slightly different timing data and thus different forward modelled IM distributions.}
              \label{IM_Range}%
    \end{figure}
    
\subsection{Parent Prior Constraint}
    
We have shown the fitted empirical IM parameter distribution, but the question still remains of how informative the priors are. Specifically, does the prior constrain the range of systematics or does it fill the space of possible systematic effects without discrimination. The range of effects that these Student's t IM distributions represents is not immediately clear from the parameter values themselves. Thus, in Fig.~\ref{IM_Range} we present the distribution of forward modeled light curve systematics sampled from the IM parent distribution for a selected target and wavelength channel. The exponential breath parameters were not modeled here so we could isolate the effects of the slope terms. The IM values were sampled from the fitted Student's t-distribution fits in Fig.~\ref{SpectrumFittedParentDistr}. From Fig.~\ref{IM_Range}, one can see that the central 68\% of instantiations is very tight and represents an arguably minimal effects model. The 95\% envelope, however, demonstrates that 27\% of the time, the effects are greater than the minimal effects model all the way up to very significant but still plausible effects.

This demonstrates one reason why the Lorentzian and more generally t-distribution proved to be a well fitting model as compared to the Gaussian. It correctly captured the minimal effects central IM changes while applying a lower but still significant probability onto the greater effects. A Gaussian on the other hand, with its comparatively thin tails has significant trouble fitting the outlier and high effects models and must thus apply a more uniform distribution across both the strong central minimal effects model and the less common maximal effects. As a result, a Gaussian overestimates the occurrence of outlier IM parameters when fitted as Fig.~\ref{SpectrumFittedParentDistr} demonstrates with the Gaussian estimated density significantly above the true empirical density in the tails.

\subsection{Per Target Spectrum Prior Analysis}

\begin{figure}
   \centering
    \includegraphics[width=\linewidth]{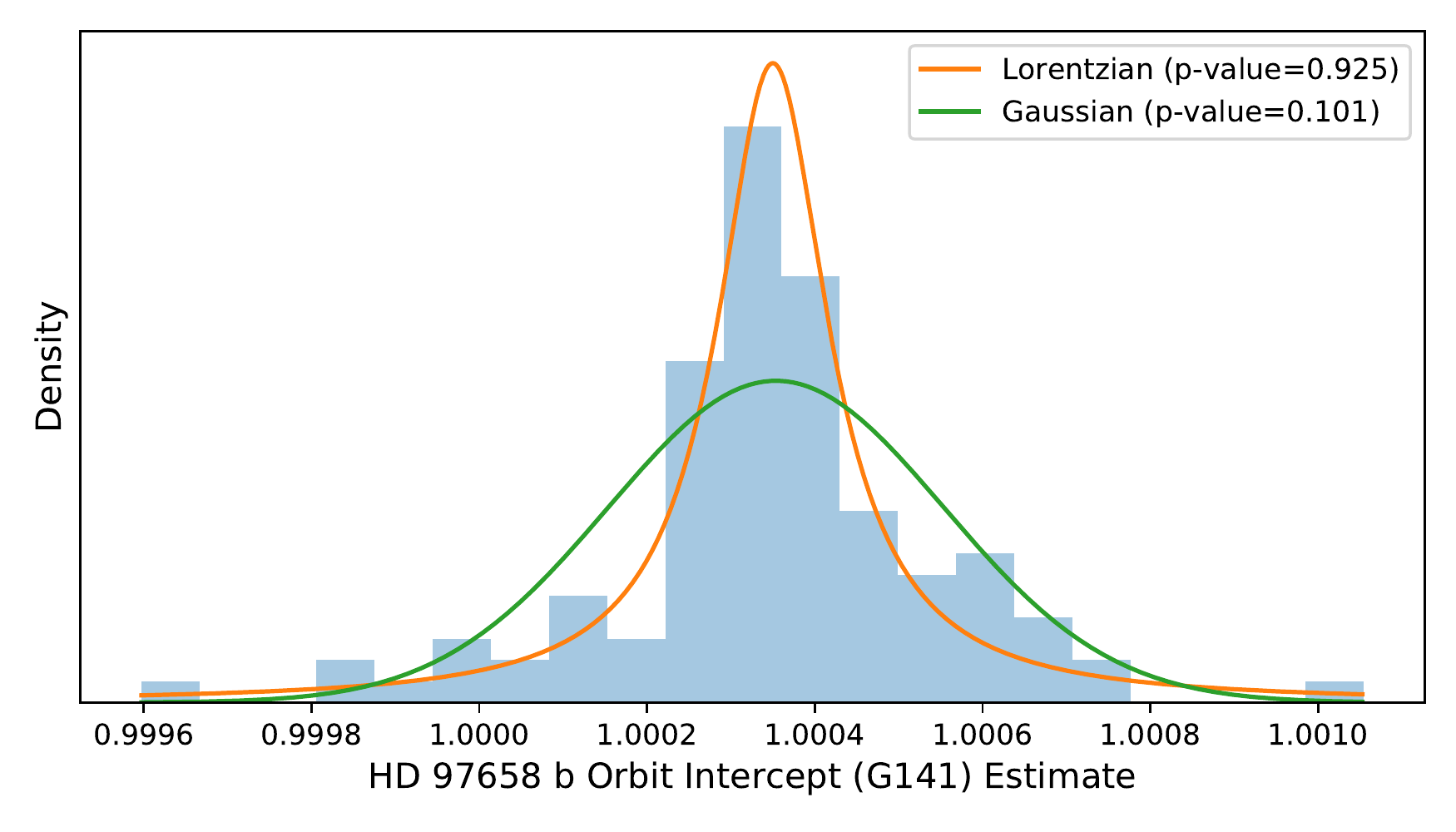}
    \includegraphics[width=\linewidth]{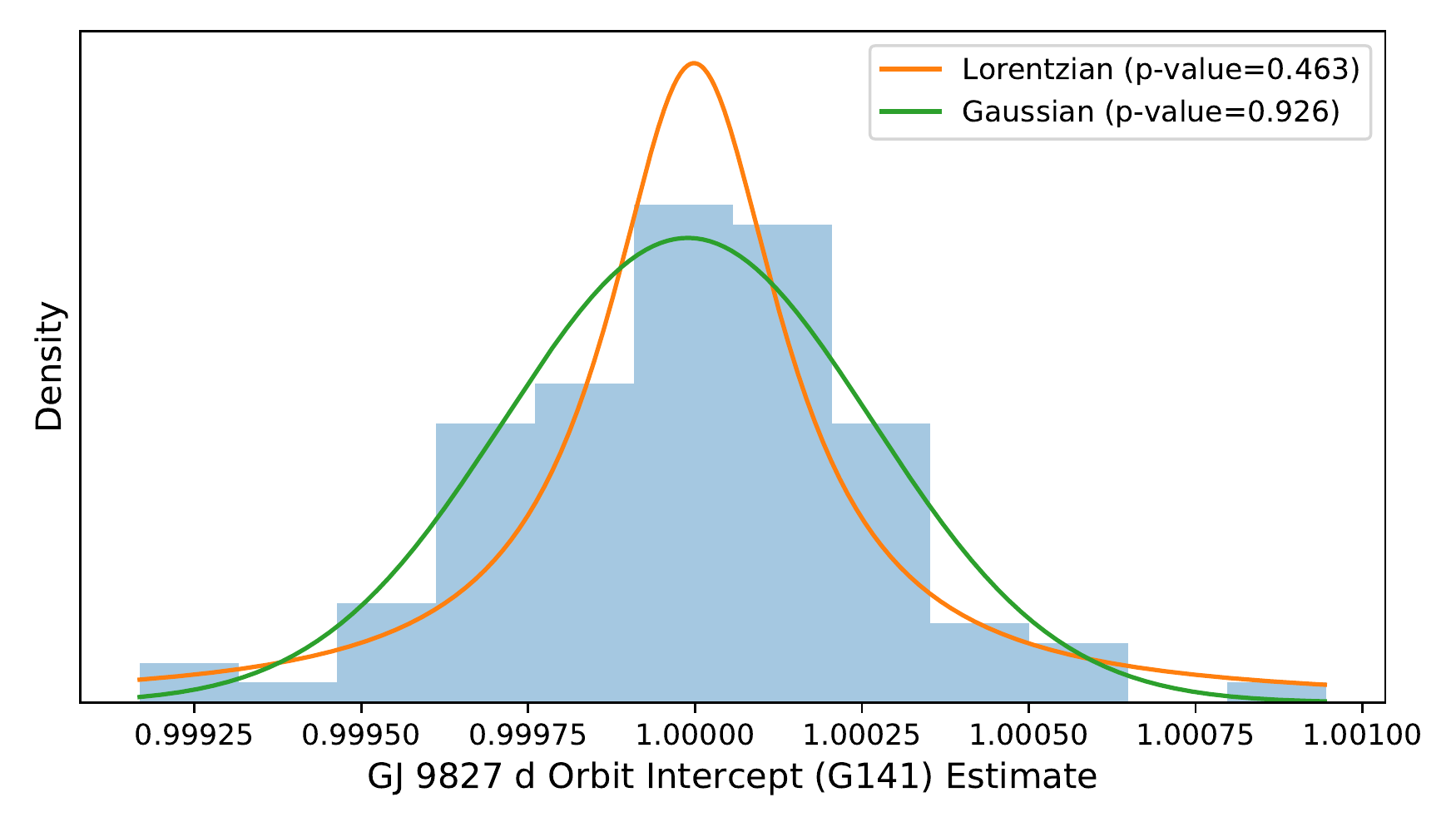}
   \caption{The orbit intercept parameter distribution for the first visits of HD 97658 b \textbf{(top)} and GJ 9827 d \textbf{(bottom)}. Fitted Lorentzian and Gaussian distributions are superimposed and the KS distributional fit test p-value is included in the legend.}
              \label{SpectrumTargetsDists}%
    \end{figure}

The focus has so far been on the Lorentzian/Student's t behaviour of the parent distribution of spectrum estimates, rather than the behavior of estimates for given targets. For the IM estimate distribution across wavelengths for individual targets, we discovered that some visits exhibited very Lorentzian shapes whereas others, while still providing a sufficient Lorentzian approximation, exhibited higher degree of freedom t-distribution behaviour nearing Gaussian behavior. The first visit of HD 87658 b orbit intercept on G141 is a key example of the Lorentzian behaviour and the first visit of GJ 97658 d orbit intercept on G141 is a key example of the more Gaussian behavior. The empirical distributions and fitted distributions for these two targets is included in Fig.~\ref{SpectrumTargetsDists}.

This range of distributional behavior adds credibility to the t-distribution being a valid parent as it agrees with our previously mentioned analysis in Fig.~\ref{T_Dist_Samples} that shows the significant range of distributional characteristics that the Student's t-distribution can exhibit. By definition, a fitted Student's t-distribution was equally or better fitting than the Gaussian or Lorentzian for all targets, as these two are special cases of the Student's t.

We speculated that the distribution of parameters for individual targets could be Gaussian. Under this hypothesis, the apparent t-distributional shape of the parent distribution over all targets could be an artifact of the varying means and standard deviations of the individual target's Gaussian distributions. Specifically, the observations could be consistent with each target visit having an IM parameter mean that is sampled from a t-distributed prior. Then, these individual targets could have a Lorentzian or Gaussian behavior around this randomly sampled t-distributed mean. This hypothesis can thus be viewed as the t-distribution being a prior on the priors of each target visit. Each target visit can have an underlying prior that is Gaussian, Lorentzian or some other distribution, but the hyper-prior over all targets and visits is this t-distributed parent prior. Without knowing the true IM distribution (i.e. prior) of a given target visit, this parent prior would thus be used when estimating these IM parameters to avoid biasing effects.

We discovered that G141 could be consistent with Gaussian priors on each target visit, and a t-distributed prior on the priors, but the STIS filters exhibited low DoF t-distributed behavior at both the target and the parent prior level. We will now analytically explore the distribution of these per visit IM parameters.

To characterize the Gaussian or Lorentzian nature of the per target IM parameter distributions, we sampled from a Gaussian distribution with the same number of points as spectral channels in the corresponding filter, and with mean and variance set to the average of the fitted parameters for targets of that filter. We then compared how often the KS test assigned a higher p-value to a fitted Lorentzian than a fitted Gaussian. This was then compared to the actual observed times that a Lorentzian fit better than a Gaussian. This test can thus indicate if the per target visit IM parameter distributions are more Gaussian or more Lorentzian in nature. The data can be seen in Table~\ref{SpectrumFitPercent}. G141 fluctuated around the value expected under a Gaussian assumption. On the other hand, both STIS filters G430L and G750L had a higher Lorentzian quality of fit significantly more times than would be expected under a Gaussian assumption. Further, for both filters the Lorentzian had a higher p-value more often than the Gaussian for all of the parameters.

\begin{table}
\caption{Percent of times the Lorentzian fit a target's spectrum IM parameter distribution better than the Gaussian did compared to the expectation under a true Gaussian}
\label{SpectrumFitPercent}      % is used to refer this table in the text
\centering                          % used for centering table
\begin{tabular}{c c c c c}        % centered columns (4 columns)
\hline\hline                 % inserts double horizontal lines
Filter & vslope & oslope & oitcp & Expected (Gaussian) \\    % table heading 
\hline                        % inserts single horizontal line
   G141 & 30.8\% & 23.1\% & 26.2\% & 27.5\%\\
%   G430L & 55.6\% & 59.3\% & 70.4\% & 31.9\%\\ % low resolution G430L
   G430L & 44.4\% & 59.3\% & 70.0\% & 28.9\%\\
   G750L & 55.2\% & 79.3\% & 79.3\% & 28.9\%\\
\hline                                   %inserts single line
\end{tabular}
\end{table}

This suggests that STIS may exhibit Lorentzian behavior on a per target basis, whereas the data for G141 on WFC3 is consistent with a t-distributed mean shifting process followed by approximately Gaussian samples around this t-distributed mean. Consistent with the whitelight IM distributions (Fig. \ref{WhitelightFittedParentDistr}), the per target means for the WFC3 filter do appear to follow an approximately Lorentzian distribution, but there is not enough samples to properly quantify exactly how close. However, for the purposes of developing an IM parent prior, a fitted t-distribution can explain the behavior of both STIS and HST if the underlying sample mean correctly follows a low degree of freedom t-distribution. This holds even if visit subpopulations follow a more Gaussian behaviour around their t-distributed mean, as such a setup would still closely follow a t-distribution when not conditioned on a specific target, thus satisfying the prior.

Further, we previously demonstrated that the initial priors for the IM parameters were wide and thus are not a strong influence in this distribution shape. This suggests, that this Lorentzian behavior of STIS originates in the data itself and the Lorentzian related behavior of G141 is also data informed rather than prior informed.

By virtue of this analysis, we can conclude that the IM appears to be characterized by a t-distributed noise source that has a variable effect on the parameter estimates. Whether, this is an artifact of estimation or a physical phenomenon is left to future study and would likely require an even larger sample size to generate demonstrable proof. The influence of different principal investigators on the derived estimates is one such area of research this paper did not address. Our IM formulation aimed to minimize the effects of how centered the transit is in the collected data, but effects of choices such as cadence at which the data was read and how long the electron wells were intended to fill in a single observation may enable explanation of additional variance. This distributional analysis part of this paper can thus serve as motivation for future analysis of this kind as sample sizes continue to increase.

\section{Implications of Prior Analysis}

Our findings have strong implications for parameter fitting procedures in low SNR regimes. Poor or incorrect choices of prior distributions can bias fit results, and our findings demonstrate that the Gaussian distribution does not closely model the empirically observed instrument model parameter characteristics. Conversely, the Student's t-distribution closely fits the observed characteristics. Further, as will be discussed in more detail, the Student's t-distribution is more resilient to outliers while not erroneously suppressing data supported parameter estimates far in the tails of the observed distribution.

\subsection{Parent Priors Versus Non-Informative Priors}

We tested the difference between models fit using the relatively non-informative priors initially discussed which gave roughly uniform and wide probability over the parameter space versus the results when using the Student's t fitted distributions as priors. The resulting light curves and spectrum where slightly different, but the change, while varying across the targets we tested, was never more than 10\% of the fitted spectrum uncertainty. This indicates that for the HST targets we tested, the fits were data informed due to the relative strength of the data signal for these sensors. However, for other regimes of signal to noise with other instruments, this approach of empirically fit prior distributions will be important. While the data informed nature of this problem reduces the importance of strong priors, this paper's discovery that a Gaussian does not adequately capture the observed instrument model distributional properties, suggests that the Student's t-distribution is an empirically supported distribution for strong parent priors robust from and accepting of data supported outliers.

\subsection{Parent Distribution Implications}

The demonstrated evidence for a low degree of freedom Student's t-distribution parent distribution both aids the fitting process and is supporting evidence for the validity of comparative analysis. It is expected that individual targets will deviate from this parent t-distribution due to specifics of how observations for a particular visit were taken. However, the quality of fit of the t-distribution suggests that there is a significant shared error origin for each of the parameters and filters. For the purposes of wide scale analysis, this supports the hypothesis that we are marginalizing the same kinds of error and thus, the dominating influence of the IM on a particular target is common to all the others. This provides supporting evidence that direct comparison of spectral features between targets is valid as they have had the same linear systematic light curve influences in the reduction.

The exact physical reasons why a Student's t-distribution fits closely have not been determined by our analysis, and are a ripe area for future study. We do, however, speculate that, as the t-distribution is seen in mean estimation and coefficient estimation processes with few supporting points such as linear regression, the underlying error source may be analogous to a mean estimation process.

While this analysis cannot directly prove that a Student's t parent distribution is the underlying physical explanation, it does show that a Gaussian is unable to model the IM parameter distributions, most notably the thicker observed tails. A Gaussian can thus either closely model the central distribution but suppress the observed tails, or model the tails but be too wide to be informative. This work demonstrates that a Student's t is able to empirically model the distribution closely without this tradeoff. Further, the Student's t is less susceptible to outliers with its wider tails, and as apparent outliers are common in this problem, the Student's t-distribution is a better candidate for directly modeling the data without arbitrarily removing apparent outlier parameters.

\section{Instrument Model Simplifications}

\subsection{Instrument Model Linearity}

\begin{figure}
   \centering
    \includegraphics[width=\linewidth]{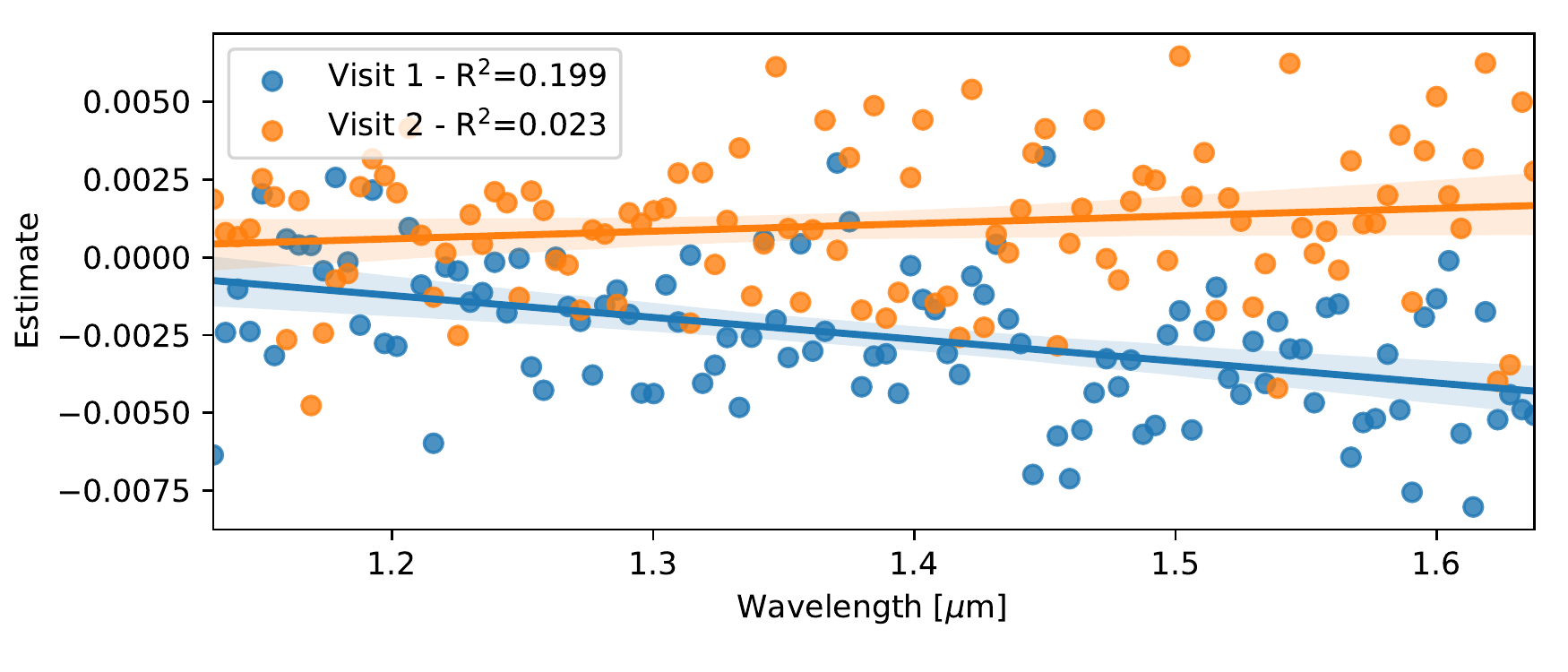}
    \includegraphics[width=\linewidth]{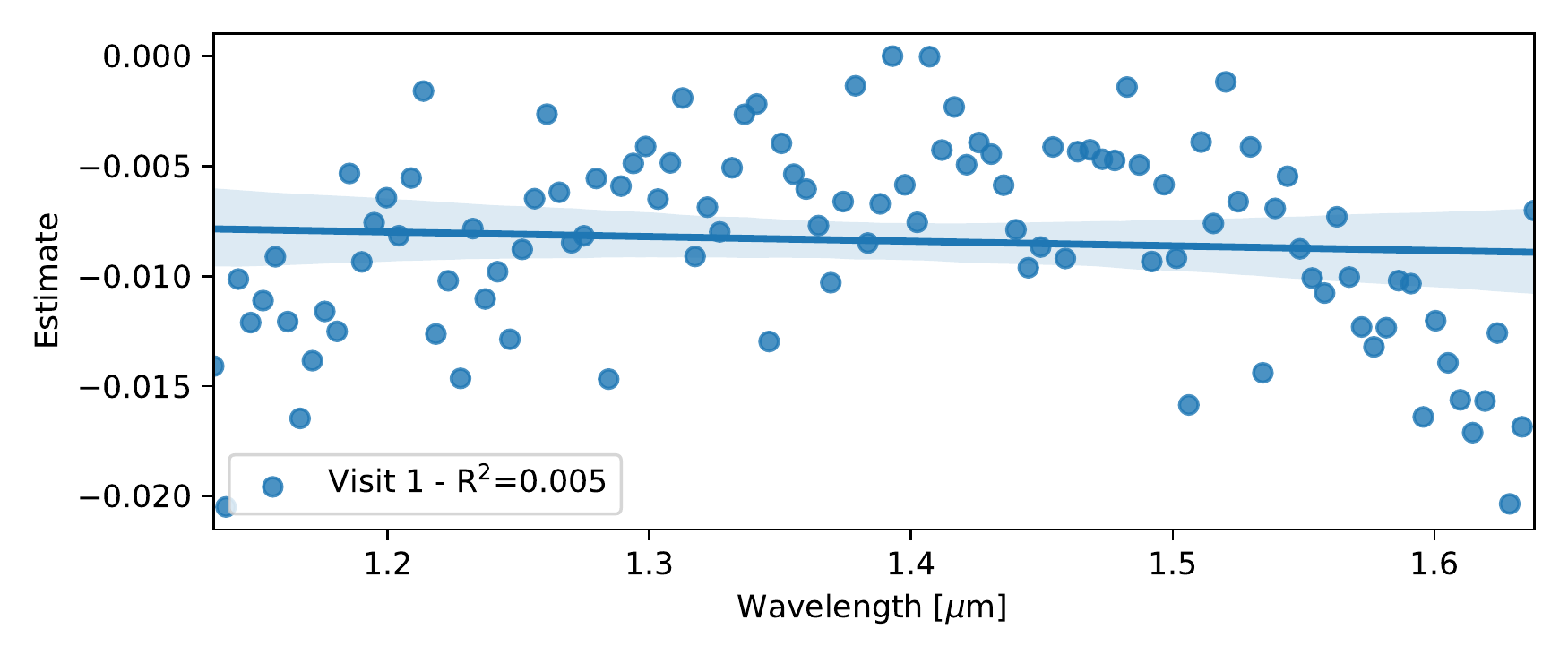}
   \caption{\textbf{Top:} Visit slope estimate distribution by wavelength for two visits of HAT-P-38 b on G141 showing one with a weak linear signal and one with a strong negative correlation to wavelength. The R$^2$ value of the visit fit is shown in the legend. \textbf{Bottom:} Visit slope estimates for HAT-P-17 b on G141. This target exemplified the behavior of targets that had non-linear parameter trends.}
              \label{InstrumentModelLinearity}%
    \end{figure}

We next explored linearity between instrument model parameter values and wavelength. We discovered that, as illustrated in Fig.~\ref{InstrumentModelLinearity}, targets such as HAT-P-38 exhibited a linear relationship between IM parameter and wavelength for at least some of the visits whereas other visits and targets did not show a significant linear relationship. Further, this reference relationship in Fig.~\ref{InstrumentModelLinearity} was not the strongest linear relationship observed but was rather selected to show a more typical relationship strength when linearity was detected. Additionally, the bottom plot in Fig.~\ref{InstrumentModelLinearity} shows one target which exhibited a strong non-linear behavior as evidenced by the upward sloping trend at the start followed by a sharp anticorrelation at the end. These linear and, more rarely, non-linear relationships were observed across G141, G430L and G750L.

To quantify this linearity, we took the fitted IM parameters and the middle wavelength value of their corresponding wavelength channel and then performed simple linear regression of the IM parameters on wavelength. We discovered that this regression had a statistically significant slope at the 95\% significance level $36.0\%$ of the time for G141 and $64.8\%$ for G750L. Further, for G750L we discovered that some of the relationships exhibited an apparent sinusoidal behaviour for some wavelength bands but the data size was not large enough to warrant demonstrative proof of this relationship.

This finding indicates that a linear IM behaviour is observed in wavelength but that further factors also play a role in this wavelength dependence so this relationship cannot be used to perfectly capture this wavelength dependent relationship. The $r^2$ values make this clear with a maximum $r^2$ of $0.71$ and a median of $0.018$ for G141, showing that in the typical case, very little variance is explained by the linearity. Comparatively, G750L had a maximum of $0.61$ and median of $0.13$ indicating that it typically showed more linearity than G141. G430L also exhibited linear behavior but a lower median $r^2$ than G750L.

\begin{figure}
   \centering
   \includegraphics[width=\linewidth]{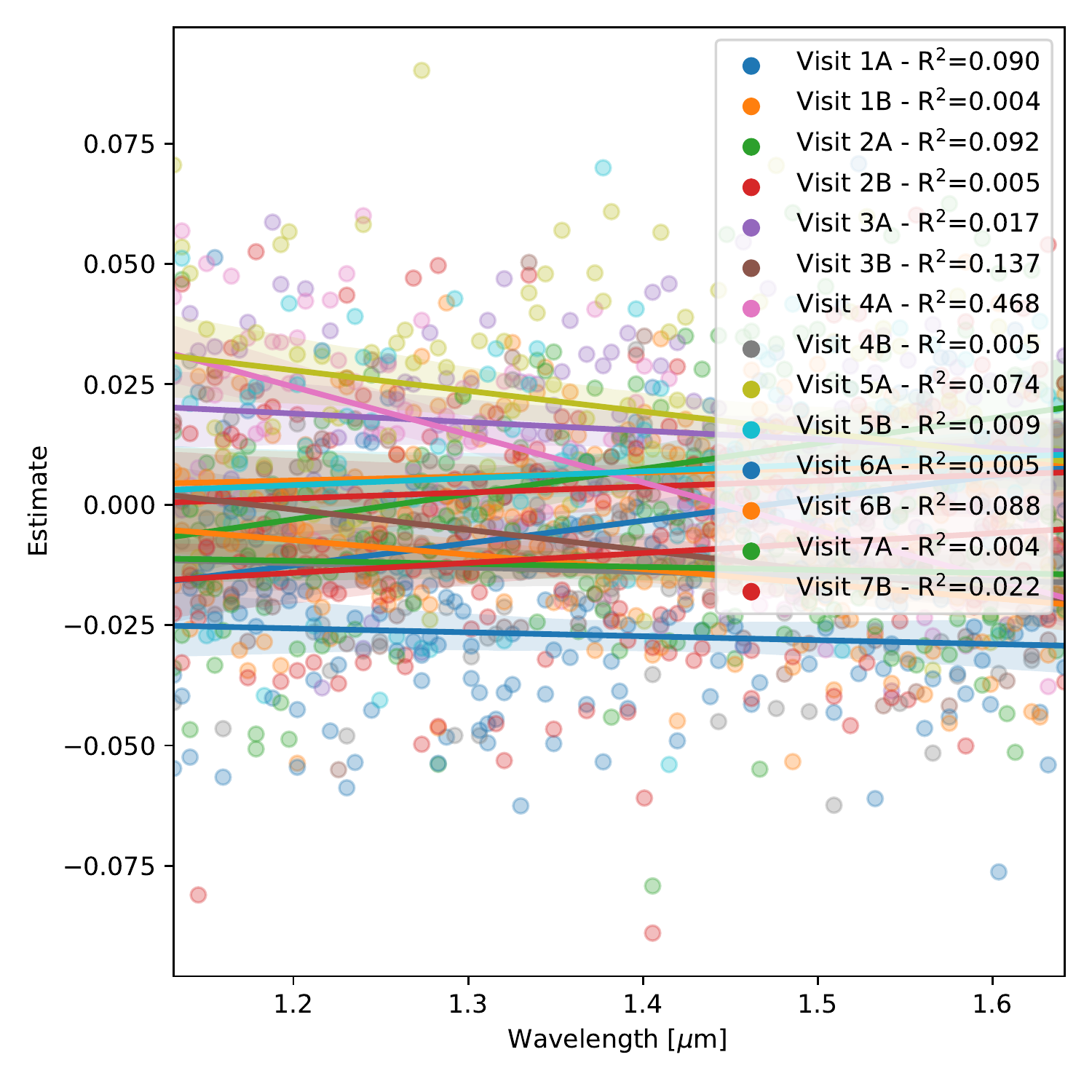}
   \caption{Orbit slope parameter distribution for GJ 1214 across the wavelengths of G141 with a plotted regression line fit between wavelengths and estimates for each visit. Legend marks the North and South scanned light curve data for the 7 visits. If the exoplanet target's characteristics was a primary influence on the IM parameters, each visit would be expected to have similar linear and mean characteristics, but the plot clearly shows that this assumption does not hold true for GJ 1214. This implies that the exoplanet characteristics are not a significant influence if any influence exists at all.}
              \label{GJ1214_Oslope_Visits}%
    \end{figure}

We tested and rejected the hypothesis that this slope could somehow be tied to the exoplanet characteristics and thus indicate a degeneracy in the IM. The top plot in Fig.~\ref{InstrumentModelLinearity} shows the typically observed behavior in which multiple visits of the same target had different IM to wavelength relationships. This was made most clear for GJ 1214 which had 7 double scanned visits, all of which showed many different linear trends with correlation, anticorrelation and no linearity visit behavior all observed across the 7 visits. Fig.~\ref{GJ1214_Oslope_Visits} shows this discovery. Thus, the evidence strongly suggests that our model is capturing the instrument systematics and not variance tied to the exoplanet characteristics. Specifically, the exoplanet is not inducing a particular wavelength slope nor is it inducing a particular average IM parameter value.

We hypothesized that the linearity might result from our particular precalibration and postcalibration processes, and could be the result of unfactored effects in wavelength correction. We were unable to discover any relationship from our calibration steps and raw data characteristics to whether a particular target would exhibit a linear or non-linear IM behaviour. We thus did not find evidence to support a link between the linearity and our calibration process, but the effects of calibration in general is a ripe area for continued exploration.
    
We tested strictly enforcing these discovered linear relationships in order to see if it was a viable simplification of the IM. Rather than having independently fit parameters for each wavelength channel, all channels IM parameters would be held to a simple linear model with two parameters, the slope and the mean with wavelength as the predictor. This drastically reduced the total IM complexity. Specifically, rather than having separate parameters for each channel, it now had just two linear parameters to capture the relationship of all channels. For most targets the impact on the resulting spectrum was minimal, but when analyzing targets with good low variance spectrum, such as HAT-P-26 b on G141, we determined that the variance of the spectrum did increase. This suggested that the variance around the linear IM relationship was valid, and did improve the spectrum. As will be further explored in Section 6 on model performance, our current model was not overfitting as measured by the marginalized light curve residual standard deviation. Thus we determined that this simplification, while interesting, was not necessary and a more complex model was justified to fully capture the variance.

\section{Model Performance}

Our model was evaluated on multiple metrics, but the most relevant metric chosen was the average light curve Residual Standard Deviation in units of Shot Noise (RSDSN). Other metrics such as reduced Chi-square were considered but as noted in \citet{andrae10}, they are not valid except for purely linear models. We will thus study our performance by the RSDSN measure and then compare this result to that of a competing model.

\begin{figure*}
   \centering
   \includegraphics[width=0.95\textwidth]{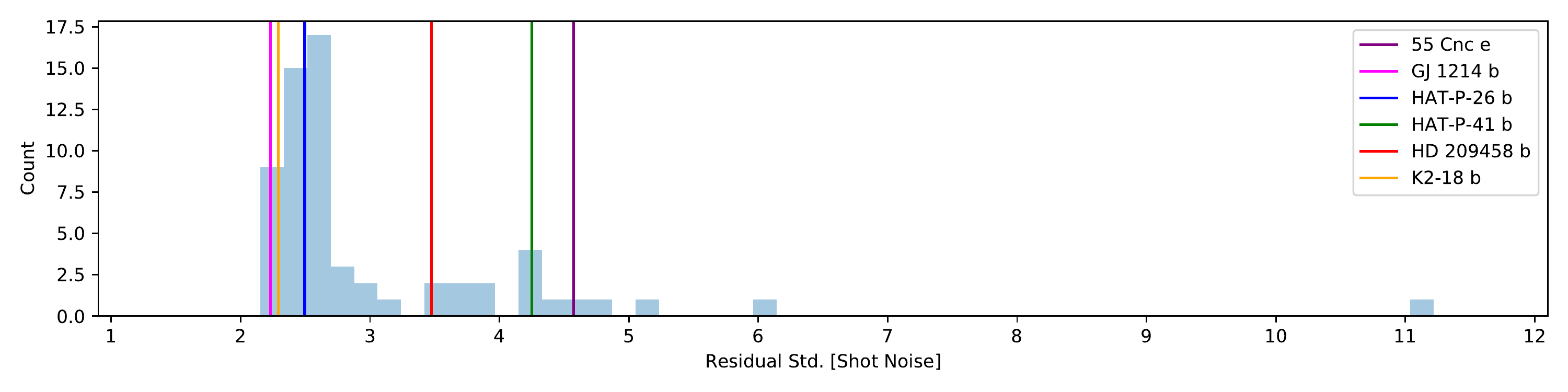}
   \caption{Histogram of average Residual Standard Deviation in units of Shot Noise (RSDSN) across targets observed by G141. $3.25$ RSDSN appears to be the cutoff between the low RSDSN grouping and the extended tail of higher RSDSN outliers.}
              \label{ResidualStdPerPhotonNoiseG141}%
    \end{figure*}

We discovered that for the significant majority of targets on G141, our model performance was under 3 times the shot noise as seen in Fig.~\ref{ResidualStdPerPhotonNoiseG141}. This shot noise limit is the theoretical minimum standard deviation so the actual light curve residual standard deviation is expected to be higher due to other factors such as irregularities in the scanning rate. When we compared our estimated uncertainties on the transmission spectra to several targets studied in the literature, we found ours agreed closely. Thus, given our RSDSN distribution and for our computation of shot noise, an RSDSN under 3 suggests good data quality. This exact value may vary depending upon how shot noise is estimated as well as how much pixel binning is done, but our threshold and our analysis should agree with peer estimations to within a scaling factor. Analyzing these thresholds and different approaches to uncertainty estimation, would be a ripe area for future study. From the figure, a RSDSN value of 3.25 was determined to be a good threshold for flagging targets which might not work with the IM, whether because of underlying data issues or because of some other formulation being needed.

We compared this metric performance for several targets to the simple model proposed by \citet{tsiaras16}. This model encapsulates less wavelength dependence and uses the whitelight IM while only fitting for $R_P/R*$, a normalization factor and a visit long slope. We discovered that it slightly increased the variance of spectrum on a selection of good targets and achieved similar residual metric performance on some targets and lower performance on others. This thus indicates that they achieved a similar class of performance, but as our performance was still well above the shot noise limit, there was not an indication that our approach led to overfitting.

The exact RSDSN multiplicative factor partially depends upon how many pixels are binned together for a wavelength channel light curve (i.e. the more pixels that are binned together the higher the RSDSN will be as will be shown in a later section). However, the relative metric difference between targets will have a similar relationship even as the common specific multiplicative factor changes if all targets have a greater or lesser pixel averaging. We observed this with G430L and G750L which had a different multiplicative factor they centered around, but otherwise exhibited the same presence of a strong grouping at the low metric values and then a long tail of outlier targets.

\subsection{STIS Linearity}

While this paper analyzed a linear model with exponential breathing parameters, non-linear IMs are sometimes used for STIS targets \citep{sing16}. The RSDSN metric provides a proxy for quality of fit and if strong non-linear IM effects are present in a signal, then the model residuals relative to the theoretical minimum---what RSDSN measures---would be elevated above those of a target with only linear effects and the same random noise profile. Thus, in the choice to include non-linear effects, the RSDSN metric could be a good approach for identifying which targets are not sufficiently marginalized by a linear IM and should thus be improved by modeling polynomial behavior.

\subsection{Interpretation of Residual Standard Deviation}

As noted, Fig.~\ref{ResidualStdPerPhotonNoiseG141} contains the distribution of the residual metric over the targets studied from G141. Although most targets performed between the 2 to 3 value range, there was a long tail of targets with light curve residuals more than 3 times the shot noise.

The figure highlights the metric value for several targets. We discovered that for good targets for which we derived clean spectra that agreed well with the literature, such as HAT-P-26 b, the metric value was low and in the range of 2 to 3 times the expected shot noise. Conversely, there were several targets for which our analysis method generated spectra with high scatter, implausible bimodality or high transit depth estimate variance, and 3 of these targets are highlighted in the figure, HD 209458 b, 55 Cnc e, and HAT-P-41 b. All three of these targets had a residual metric greater than 3.

This evidence suggests that this residual metric can be used to detect which exoplanet targets achieve results that agree well with the literature and which do not achieve repeatable results. Manually reviewing the light curve residuals of high residual metric targets did not indicate remaining systematics, but exploration of higher dimensionality models could be a fruitful area of exploration for the high RSDSN targets.

This suggested that the source of the low quality spectrum derived from this method for the bad targets was either in the data itself or our data reduction process. Although either could be the source of the problem, the metric value allows us to quickly determine if we can trust a particular spectrum, at least for the targets in this study.

\subsection{Per Channel Performance}

\begin{figure}
   \centering
   \includegraphics[width=\linewidth]{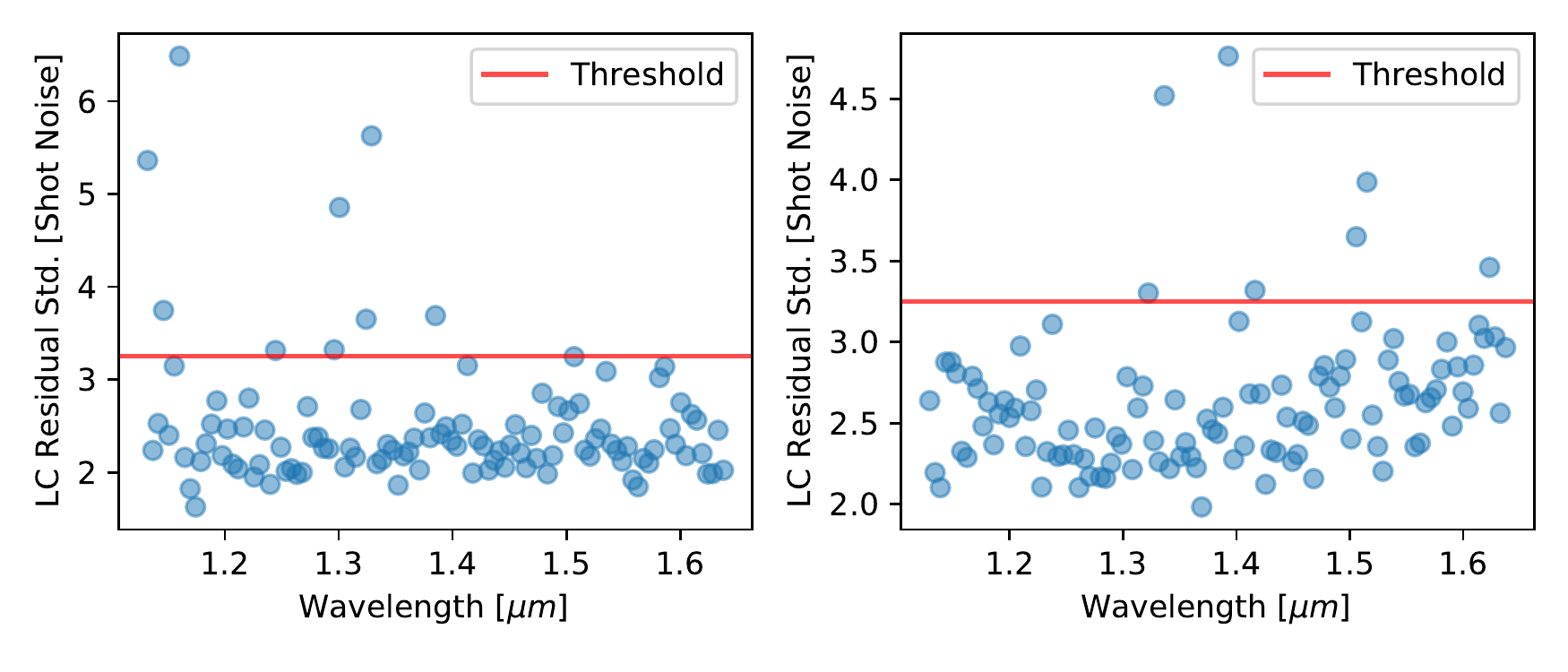}
   \caption{The residual standard deviation in units of shot noise for the light curve of each fitted wavelength channel for the two targets displayed (\textbf{left}: HAT-P-26 b, \textbf{right}: GJ 3470 b, both in G141). Overlays the 3.25 RSDSN fit quality threshold.}
              \label{RSDMSpectrum}%
    \end{figure}

The above metric computation required computing the metric separately across each wavelength channel and then averaging each of these separately computed metric values. We now focus on the metric value for each channel. We discovered that nearly every target exhibited a strong central mean with significant outliers occurring at several points along the spectrum. Fig.~\ref{RSDMSpectrum} demonstrates this behaviour, and places the previously determined 3.25 RSDSN threshold line to give a concrete sense of how poor quality channels could be flagged.

Further, we discovered that some targets exhibited an apparent linear or polynomial metric relationship across wavelength as GJ 3470 b in the figure shows. The underlying explanation of this is unknown, but it does provide evidence for a wavelength dependent light curve quality.

The existence of outlier fit qualities and wavelength fit quality trends suggests that, for atmospheric analysis, this per wavelength metric should be compared to the derived atmospheric spectrum to ensure that poor fit qualities at specific wavelengths or systematic light curve degradation are not the cause of any spectral features.

\subsection{Poor Quality Wavelength Channels}

\begin{figure}
   \centering
    \includegraphics[width=\linewidth]{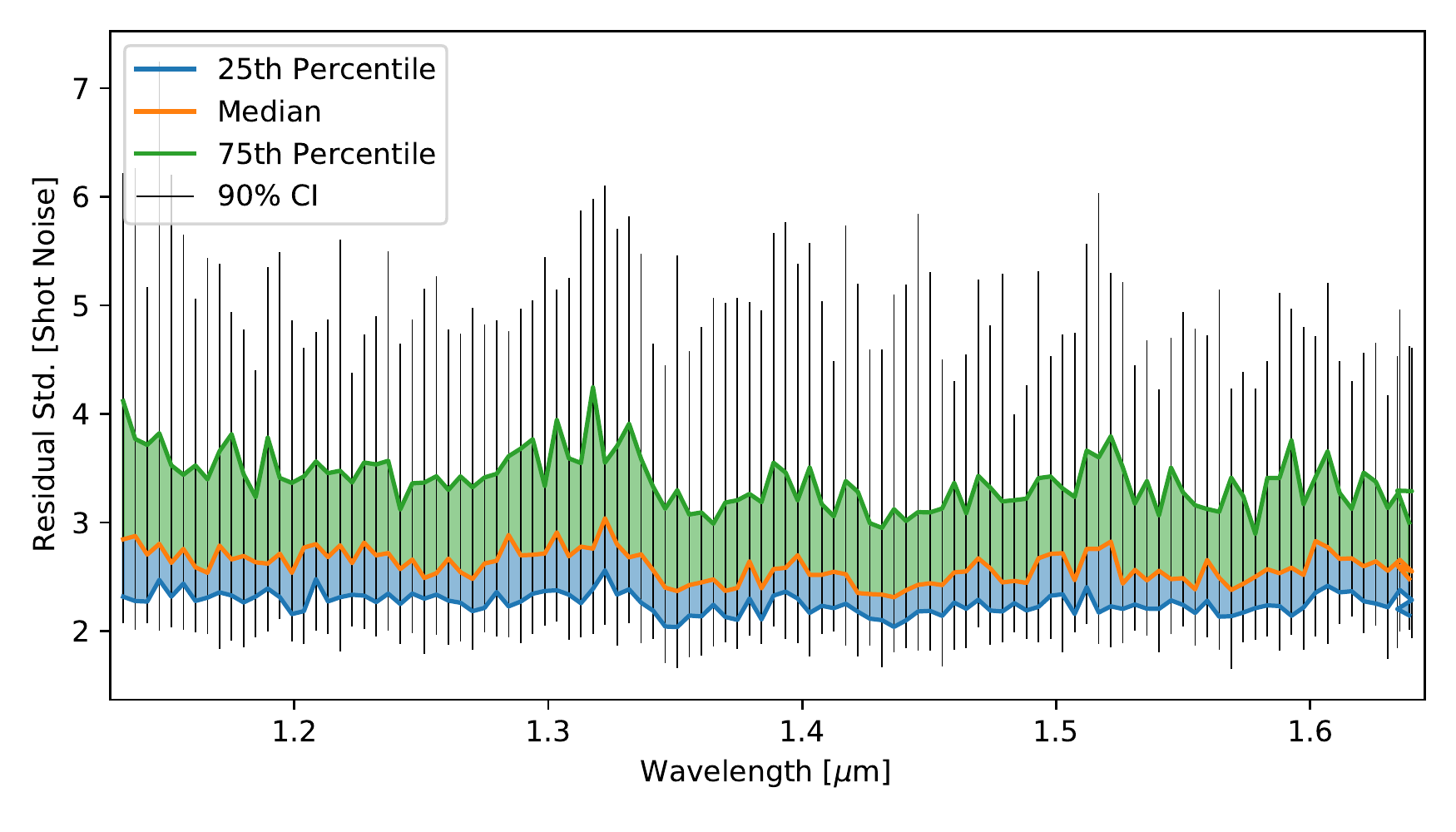}
    \includegraphics[width=\linewidth]{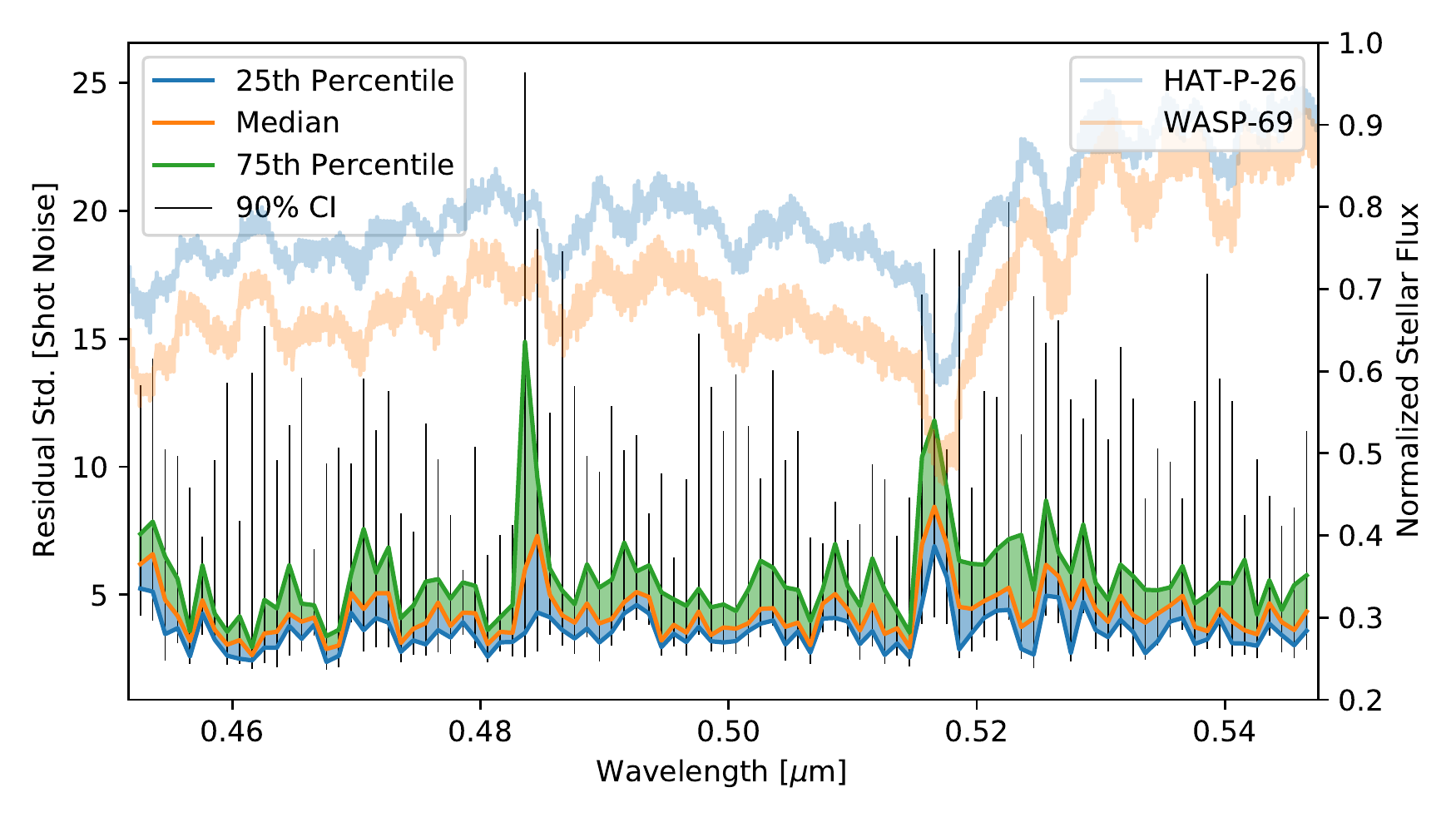}
    \includegraphics[width=\linewidth]{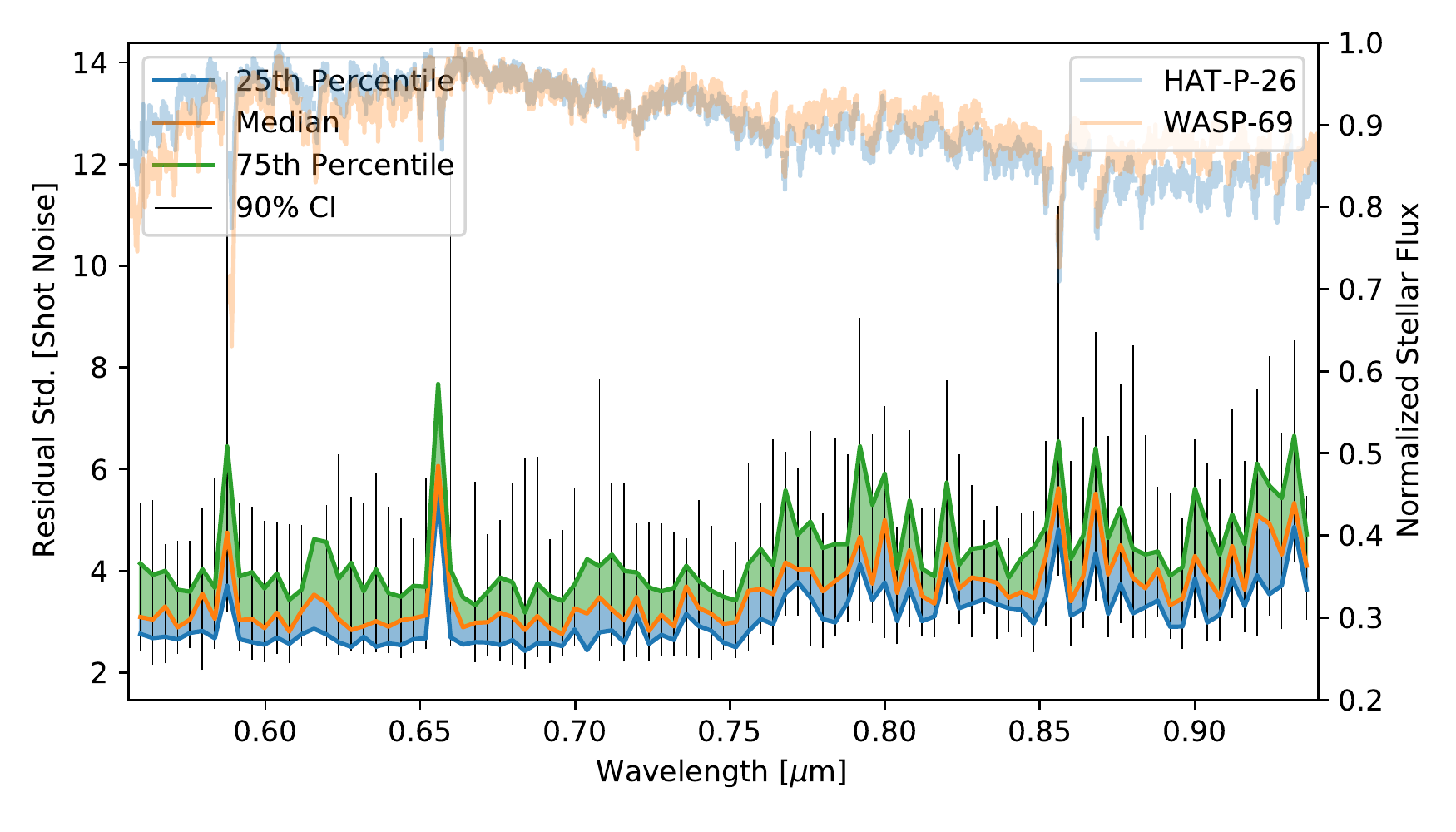}
   \caption{The distribution of the light curve residual standard deviation metric per wavelength for the the 3 filters (\textbf{top}: G141; \textbf{middle}: G430L; \textbf{bottom}: G750L). Left axis scales differ between the three filters. The stellar spectra for HAT-P-26 and WASP-69 are plotted for G430L and G750L to visualize correlation between stellar features and residual features.}
              \label{RSDMByWavelength}%
    \end{figure}

Continuing the residual metric analysis, we attempted to determine if there were any wavelength channels for which the 3 filters gave consistently poor quality light curves. This was done by binning the RSDSN of each targets wavelength channels by wavelength. The derived distribution can be seen at Fig.~\ref{RSDMByWavelength}.

We discovered that G141 did not show an indication of consistently bad wavelength channels, whereas G430L and G750L on STIS had several wavelength channels that gave consistently high residual light curves. For G430L, the channels centered on 484.6 nm and 516.6 nm had consistently anomalously high RSDSN. Analogously, on the bluer side of the G750L spectrum, 587.7 nm and 655.8 nm had anomalous RSDSN. Fig.~\ref{RSDMByWavelength} overlays the stellar spectra for two targets, HAT-P-26 and WASP-69, over the per wavelength RSDSN for G430L and G750L. A visual analysis demonstrates that these high residual channels are correlated with stellar absorption features. Future analysis of these channels could explore whether additional effects beyond the decreased SNR, such as potential shortcomings of interpolation over these wavelengths in the initial data reduction, could be a factor in the prominence of these higher residual wavelength channels.

Further, for G750L we discovered that higher wavelengths than 750 nm had elevated variance compared to the bluer wavelengths and exhibited significantly more RSDSN fluctuation between channels. We hypothesize that this elevated variance on the redder side is explained by two factors. First, the throughput of G750L at wavelengths $>800$ nm decays by ~2\% compared to the bluer side. The second factor is the existence of residual fringes in our reduction. There are some significant fringing effects that affect the red side of the G750L grism \citep{Nikolov14, Loth18}. To remove the fringes, we use a contemporaneous flat fringe that was taken during the observation of each target. However, the contemporaneous flat fringe does not fully reproduce the fringes in the redder side of the spectrum. Thus, when we divide the spectrum by the flat fringe model, there will still be some residuals fringes that will elevate the variance.

We conclude from this spectrum residual analysis that over a nearly 30 source catalogue, G750L and G430L have several channels that are consistently and systematically problematic, although this is largely explained by stellar absorption features. These high variance channels should be noted and considered in spectrum analysis as spectral features derived from them come from a separate distribution of light curve errors. Thus, these channels could potentially possess different light curve behavior which could impact the resolved transit depth estimates.

\subsection{Implications for Spectral Averaging}

Pixel averaging well beyond the dispersion limited minimum binning is common to try to reduce noise. Our analysis has shown that some wavelength channels exhibit high residual light curves compared to the rest. A potential benefit of using the maximum spectral resolution is that these ill performing channels can be identified and could be flagged or removed. It is possible that spectral averaging with these poor channels could bias the transit depth estimates for that spectral range, and by using the full spectral resolution, this biasing could be avoided or otherwise controlled for. The potential biasing impact is a ripe area for future research.

\section{Conclusion}

We have demonstrated that a simple IM works across 2 instruments and 3 filters as evidenced by the residual variance metric. Further, we have demonstrated the existence of an empirically supported parent distribution for IM parameters that closely follows a low degree of freedom Student's t fit and shown that the distribution is non-Gaussian. The IM parameters are not expected to strictly follow this parent due to observational differences between targets, but the utility of this parent distribution is that it better constrains the kind and distribution of errors observed. As seen in G141, individual targets may even exhibit Gaussian distributions of their parameters. Thus, a prior on that given target would be Gaussian, but the prior on the prior, what we refer to as the parent prior, empirically follows a low DoF t-distribution closely. This prior on the prior thus provides a general non-biasing prior estimate that can be applied across targets. A wider implication of the t-distributed behavior we observed is that, for empirically fit priors, a t-distribution may be a better choice than a Gaussian for a robust and tight fit. We conclude this because thick tails and outliers appear to be a potential common feature of parameter estimates, features which the Student's t-distribution fits significantly better than the Gaussian.

Further, the discovery of the consistent empirical parent distributions, with significant differences only in the mean and variance of these fitted distributions, adds credibility to wide scale analysis based off of the resolved spectrum. Specifically, it suggests that the marginalized error by the parameters has a common origin and thus that derived spectra have the same error removed and any spectral feature influence of this IM is common to all targets. This supports direct comparison across resolved spectrum. Additionally, we have demonstrated that this t-distribution parent prior can be empirically fit. This allows a general approach to fitting against lower quality data, and, in this case, avoids potentially biasing the results with a wrong assumption of the prior, such as assuming it is a Gaussian.

Our analysis also determined the existence of a linear wavelength IM dependence, but showed that this typically explained only a small fraction of the total variance. Conversely, we discovered that a simple estimation strategy for the exponential breathing parameters based solely on timing data achieved close fit quality compared to a $\chi^2$ parameter fit. However, fitting by $\chi^2$ did not appear to be overfitting and was thus still justified.

Additionally, we discovered that the marginalized light curve residual variance can be used as a method to identify which targets will not reproduce results consistent with the surrounding literature and can thus be used to temper confidence in any resolved features or suggest the use of a different IM.

Thus, we have demonstrated that a simple IM possesses well constrained behaviour across a range of instruments and filters. Finally, we have provided a simple metric to determine where our generic model provides a high quality fit and where more investigation and analysis may be important.

\subsection{Future Work}

This work would not be possible without wide scale analysis across many targets and multiple instruments. Thus, a key future direction is to extend this analysis to even more instruments, filters and targets to identify if and how the derived relationships hold in additional contexts and as the sample size increases even more. Finally, future work can explore causal reasons why the Student's t-distribution conforms so well to the data.

\subsection{Application to Comparative Planetology}

A key motivation for a robust IM is the application to comparative planetology, and this IM can provide uniform estimation of spectrum across a large number of targets. A large scale analysis using the same IM and estimation methodology presented in this paper is performed in \citet{roudier2021}. Analyses like this referenced one, can benefit from applying the IM analysis techniques employed in this paper, and specific metrics such as RSDSN can be used to flag targets in the set that might not conform to the IM assumptions.

\section{Acknowledgements}

This paper was made possible by the constant support of the co-authors Mark Swain, Gael Roudier and Raissa Estrela as well as the infrastructure and computing resources developed and managed by Albert Niessner. The rest of the team at JPL working with the EXCALIBUR pipeline were instrumental in validating conclusions and providing insight into potential explanations. Further, the support of the JPL Education Office and Caltech SFP was instrumental in making the research possible. Final thanks go to the HST operators and team as well as NExScI and the MAST HST archive, whose systems and data enabled this paper's analysis.

\appendix

\section{Target Details} \label{sec:target-details}

Details of each target studied, including the filters/instrument used and the program IDs is included in Table \ref{table:TargetDetails}.

\begin{longtable}[c]{c c c c}
\caption{Details for each target studied, including the HST program ID the data was collected from.\label{table:TargetDetails}} \\ 
\hline                 % inserts double horizontal lines
Target & Planet(s) & Filter(s) & Program ID(s) \\% table heading 
\hline                        % inserts single horizontal line
\endfirsthead
55 Cnc & e & G750L, G141 & 13665, 14453, 16442, 15383, 13776 \\
GJ 1132 & b & G141 & 14758 \\
GJ 1214 & b & G141 & 12325, 13021, 12251, 15400 \\
GJ 3053 & b & G141 & 14888 \\
GJ 3470 & b & G141, G750L & 13665, 13064 \\
GJ 436 & b & G141, G750L & 13665, 11622, 13338 \\
GJ 9827 & d & G141 & 15428, 15333 \\
HAT-P-1 & b & G750L, G141, G430L & 11617, 12473 \\
HAT-P-11 & b & G750L, G141, G430L & 12449, 14767 \\
HAT-P-12 & b & G750L, G141, G430L & 12181, 14260, 12473 \\
HAT-P-17 & b & G750L, G141 & 12956 \\
HAT-P-18 & b & G750L, G141, G430L & 14260, 14099 \\
HAT-P-26 & b & G750L, G141, G430L & 14260, 16166, 14110, 14767 \\
HAT-P-3 & b & G141 & 14260 \\
HAT-P-32 & b & G750L, G141, G430L & 14260, 14767 \\
HAT-P-38 & b & G141 & 14260 \\
HAT-P-41 & b & G750L, G141, G430L & 14767 \\
HATS-7 & b & G141 & 14260 \\
HD 149026 & b & G141 & 14260 \\
HD 189733 & b & G141, G430L & 12920, 12881, 11673, 11740, 13006, 14797, 12181 \\
HD 209458 & b & G750L, G141, G430L & 7508, 13467, 9064, 9447, 15813, 14797, 12181, 9055, 13776, 10081 \\
HD 97658 & b & G750L, G141 & 13501, 13665 \\
K2-18 & b & G141 & 13665, 14682 \\
K2-24 & b & G141 & 14455 \\
K2-3 & c, d & G141 & 14682 \\
K2-93 & b & G141 & 15333, 16267 \\
K2-96 & c & G141 & 15333 \\
KELT-1 & b & G141 & 14664 \\
KELT-11 & b & G141 & 15255 \\
KELT-7 & b & G750L, G141, G430L & 14767 \\
Kepler-138 & d & G141 & 13665 \\
Kepler-16 & b & G141 & 14927 \\
TRAPPIST-1 & f, e, g, d, b, c & G141 & 14500, 14873, 15304 \\
WASP-101 & b & G750L, G141, G430L & 14767 \\
WASP-103 & b & G141 & 14050, 13660 \\
WASP-107 & b & G141 & 14915 \\
WASP-12 & b & G141, G430L & 13467, 12473, 14797, 12181, 12230 \\
WASP-121 & b & G141, G750L, G430L & 14468, 15134, 14767 \\
WASP-17 & b & G141, G430L & 12181, 12473, 16166, 14918 \\
WASP-18 & b & G141 & 12181, 13467 \\
WASP-19 & b & G750L, G141, G430L & 12181, 13431, 12473 \\
WASP-29 & b & G750L, G141, G430L & 14260, 14767 \\
WASP-31 & b & G750L, G141, G430L & 12473 \\
WASP-39 & b & G750L, G141, G430L & 14260, 12473 \\
WASP-43 & b & G141 & 13467 \\
WASP-52 & b & G750L, G141, G430L & 14260, 14767 \\
WASP-6 & b & G750L, G141, G430L & 12473, 14767 \\
WASP-62 & b & G750L, G141, G430L & 14767 \\
WASP-63 & b & G141 & 14642 \\
WASP-67 & b & G141 & 14260 \\
WASP-69 & b & G750L, G141, G430L & 14260, 14767 \\
WASP-74 & b & G750L, G141, G430L & 14767 \\
WASP-76 & b & G750L, G141, G430L & 14260, 14767 \\
WASP-79 & b & G750L, G141, G430L & 14767 \\
WASP-80 & b & G750L, G141, G430L & 15131, 14260, 14767 \\
XO-1 & b & G141 & 12181 \\
XO-2 & b & G141 & 13653 \\
\hline                                   %inserts single line
\end{longtable}


\bibliography{paper}{}
\bibliographystyle{aasjournal}



\end{document}